\documentclass[article,12pt]{elsarticle}

\usepackage{graphicx}
\usepackage{siunitx}
\usepackage{lineno,hyperref}
\usepackage{xurl}
\usepackage{amsthm}  
\usepackage{rotating}  
\usepackage{amsmath}
\usepackage{amsfonts}
\usepackage{amssymb}
\usepackage{comment}
\usepackage{algorithm}%
\usepackage{graphicx}
\usepackage{algpseudocode}
\usepackage{hyperref} 
\usepackage{soul}
\usepackage{mathrsfs}
\usepackage{enumitem}
\usepackage{booktabs} 
\usepackage{tabularx} 

\usepackage[flushleft]{threeparttable}

\newtheorem{theorem}{Theorem}
\theoremstyle{definition}
\newtheorem{remark}{Remark}
\newtheorem{definition}{Definition}

\usepackage[latin1]{inputenc}

\DeclareMathOperator{\Up}{Up}
\DeclareMathOperator{\Down}{Down}
\DeclareMathOperator{\Neigh}{Neigh}

\usepackage{color}

\journal{ArXiv}

\begin{document}

\begin{frontmatter}



\title{Topological reconstruction of sampled surfaces via Morse theory}


\author[mymainaddress,mysecondaryaddress]{Franco Coltraro\corref{mycorrespondingauthor}}
\cortext[mycorrespondingauthor]{Corresponding author}
\ead{franco.coltraro@upc.edu}

\author[mysecondaryaddress,mythirdaddress]{Jaume Amor\'os}
\author[mymainaddress,mysecondaryaddress]{Maria Alberich-Carrami\~nana}
\author[mymainaddress]{Carme Torras}

\address[mymainaddress]{Institut de Rob\`otica i Inform\`atica Industrial, CSIC-UPC, \\
C/ Llorens i Artigas 4-6, 08028, Barcelona, Spain.}
\address[mysecondaryaddress]{Departament de Matem\`atiques, Universitat Polit\`ecnica de Catalunya, Barcelona, Spain.}
\address[mythirdaddress]{Centre de Recerca Matem\`atica, Barcelona, Spain}

\begin{abstract}


In this work, we study the perception problem for sampled surfaces (possibly with boundary) using tools from computational topology, specifically, how to identify their underlying topology starting from point-cloud samples in space, such as those obtained with 3D scanners. We present a reconstruction algorithm based on a careful topological study of the point sample that allows us to obtain a cellular decomposition of it using a Morse function. No triangulation or local implicit equations are used as intermediate steps, avoiding in this way reconstruction-induced artifices. The algorithm can be run without any prior knowledge of the surface topology, density or regularity of the point-sample. The results consist of a piece-wise decomposition of the given surface as a union of Morse cells (i.e. topological disks), suitable for tasks such as mesh-independent reparametrization or noise-filtering, and a small-rank cellular complex determining the topology of the surface. The algorithm, which we test with several real and synthetic surfaces, can be applied to smooth surfaces with or without boundary, embedded in an ambient space of any dimension.

\end{abstract}




\end{frontmatter}



\section{Introduction}

The domestic manipulation of cloth with robots is an increasingly relevant problem because of the everywhere presence of textiles in human environments, with promising applications ranging from dressing disabled people to automatic bed-making [\cite{Doumanoglou2016,Garcia-Camacho2020}]. Nowadays, it is relatively easy to obtain point-samples of an arbitrary and unknown garment that is presented before the robot, e.g. through the use of depth-cameras or 3D-scanners. Nevertheless, these points will have no known structure. Thus, if one wishes that the robot manipulate the garment, \textit{reconstructing} and \textit{recognizing} it from the point-cloud (i.e. to 
deduce its topology) becomes of great importance [\cite{yin2021modeling}].

\smallskip

Arguably, the main challenge faced in the automated manipulation of cloth is the high number of deformation states that textiles can present [\cite{Corrales2018}]. In contrast to rigid body manipulation, where the dynamics of the manipulated object are very well understood [\cite{Taylor:2005:CM}], there is not one single physical model that can be considered best in terms of describing the dynamics of real textiles [\cite{Nealen:2006:PDM}]. In any case, physical models of cloth behavior remain useful for developing planning and control strategies [\cite{Li2015,Colome2018}]; as well as for generating the massive data required to train learning algorithms before their deployment and tuning in the real world [\cite{jangir2020dynamic,Colome2020}]. For all of these tasks it is crucial to have an accurate and fast-to-compute reconstruction of the garment to be controlled/simulated.

\smallskip

Naturally, because of the previous reasons, the reconstruction of a surface in Euclidean space from a point-sample on it is a question to which considerable attention has been given in the areas of Computer Graphics and Computational Geometry (see [\cite{Dey2006CurveAS}] for algorithms with mathematical guarantees and [\cite{huang2022surface}] for a survey of state-of-the-art methods). Common algorithms involve triangulating the cloud points, fitting local implicit functions or more recently applying learning (i.e. Neural Networks) methods. Nevertheless, to our knowledge almost all these algorithms disregard a direct topological study of the point-cloud. Moreover, most of them focus on reconstructing \textit{watertight} surfaces (i.e. without boundary). Applying this kind of algorithms to point-clouds coming from garments can lead to incorrect results since textiles can be naturally realized as surfaces \textit{with} boundary, which is coincidentally what the majority of physical models assume in order to simulate them. One of the reasons for this focus on reconstructing surfaces without boundary may be the challenge associated in detecting boundaries of point-cloud surfaces (see [\cite{MINEO201981}]): the problem is in general ill-posed since regions of the cloud with low density could be mistaken for { gaps, with boundaries in the underlying surface}.

\smallskip

In this work, we introduce a new reconstruction algorithm that directly processes a point cloud to generate a cellular decomposition of the sampled surface. Our method produces a global piece-wise decomposition of the surface, consisting of a small number of parametrized, {explicitly} contractible, pieces. The algorithm does not rely on an intermediate triangulation or local implicit equations, thus avoiding any possible artifacts induced by such reconstructions. The algorithm is robust, consistently preserving the topological features of the sampled surface, provided these features { have a size exceeding a threshold, which can be set as low as the average distance between sample points}. From this cellular decomposition, the surface topology can be easily determined. { For instance, the cellular decomposition furnishes closed loops in any homology class having length which has the order of magnitude of the shortest possible loop in the class.} To obtain this cellular decomposition, we present for the first time in literature (to our knowledge) an algorithm to determine how (discretized) Morse cells attach to each order (see Figure \ref{esfera_morse}). Furthermore, for the case of surfaces with boundary, we develop a novel graph-theoretical method to determine robustly boundary points of the cloud. 

\smallskip

Our decomposition algorithm was first sketched by the authors in two proceedings papers [\cite{MorseEACA,MorseCEIG2}]. Here we expand greatly on that work to give full explanations on how to treat the case of surfaces with boundary, we explain for the first time in detail the theory behind our method, we give a full algorithm on how to compute the Morse cells and their attachment maps, we discuss and present results on the problem of how to parameterize the cell decomposition by flat patches and we reconstruct novel challenging point-clouds of surfaces with boundary.

\subsection{Organization}

The remainder of this paper is organized as follows: in Section \ref{sec:related} we review literature from computational and differential topology related to our method; then in Section \ref{sec_morse_teoria} we explain and develop the theoretical machinery --coming from smooth Morse theory-- needed to apply our algorithm successfully. In Section \ref{sec_local} we initiate the study of the point-cloud, giving it local structure by finding neighbors for each point, their tangent planes and the boundary curves. Section \ref{sec_morse_celdas} deals with the computation of the Morse flow, critical points, Morse cells and how they attach to each other, and it is one of the main contributions of this work. Finally, in Section \ref{sec:param2cells} we explain how to parametrize the $2$-cells by flat regions of the plane; and in Section \ref{sec_results} we present the reconstruction of five challenging point-clouds (four with boundary, one of them being a real 3D scan of a garment) using the presented algorithm.

\section{Related work}\label{sec:related}

\begin{figure}[h] \centering \includegraphics[scale=.7]{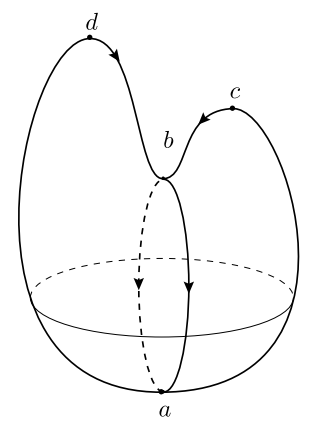} \caption{Critical points of the Morse-Smale function $f(x,y,z) = z$ on an example surface. The cellular decomposition is achieved by attaching two $2$-cells containing $d$ and $c$ (the maxima) along the curve passing through $b$ (the saddle point) and $a$ (the minimum). \label{esfera_morse}} \end{figure}

Since its inception, differential topologists have focused on the problem of parametrizing manifolds through the use of Morse functions. A smooth map $f : M \to \mathbb{R}$ defined on a compact manifold without boundary is called {\em Morse} if it has only finitely many critical points, and at each of these points, the Hessian $H(f)$ is non-degenerate. Classical Morse theory (see [\cite{Hirsch1976DiffTopo}]) demonstrates that a generic Morse function $f$ induces two distinct decompositions of the manifold $M$ via its gradient flow:
	
\begin{enumerate} 
\item {\em As a CW complex} (see [\cite{Munkres1984ElementsOA}]): Each critical point of $f$, along with its unstable manifold for the vector field $-\nabla f$, defines a cell that is topologically a ball, whose boundary attaches to lower-dimensional cells (see Figure \ref{esfera_morse}). This results in a global piece-wise parametrization of $M$, and a Morse-Smale complex, where the critical points of $f$ serve as a basis, yielding the singular homology of $M$.
	
\smallskip \item {\em As level sets}: The manifold $M$ is foliated by the level sets $f^{-1}(c)$. For regular values of $c$, these level sets are submanifolds of $M$ with codimension 1, and a diffeomorphism $f^{-1}(c_1) \cong f^{-1}(c_2)$ exists if no critical values of $f$ fall between $c_1$ and $c_2$. When $c$ crosses a critical value of $f$, the level set undergoes a surgery (see [\cite{Hirsch1976DiffTopo}] and Section \ref{sec_morse_teoria}). 
\end{enumerate}

The strength of Morse theory lies in the fact that Morse functions, along with the Morse-Smale transversality conditions necessary for this analysis, are generic among $\mathcal{C}^2$ maps from $M$ to $\mathbb{R}$. For example, a height function in a random direction in $\mathbb{R}^N$ is almost certainly a Morse-Smale function, with the probability that it is not being zero. Morse theory also generalizes to manifolds with boundaries via stratified spaces [\cite{Goresky1988StratifiedMT}], as discussed in detail in this work. \smallskip

The application of Morse theory directly to the sample point-cloud of a surface $S$ was first suggested by [\cite{Gao2008MorseSmaleD,Zhu2009TopologicalDP}], who proposed an algorithm for point-clouds with a known, uniform sampling density. Later, [\cite{Cazals2013TowardsMT}] proposed a Morse decomposition method for point-clouds sampling manifolds without boundary in any dimension. However, these approaches do not fully address issues such as cell parametrization or attachment maps, which are crucial in robotic applications where point-clouds of textiles may need to be filtered and down-sampled for simulation or control, as outlined in the introduction. We build on the gradient flows of [\cite{Gao2008MorseSmaleD,Zhu2009TopologicalDP}], but propose a novel method for identifying critical points and their Morse cells by examining level sections of these flows.

\section{Preliminaries: smooth Morse theory}\label{sec_morse_teoria}

In this section we present a summary of smooth Morse theory for surfaces with and without boundary. We first explain briefly the case without boundary, which encapsulates the main ideas of the field. For a summary of this part, including what is needed to run our algorithm see Section \ref{summary_complex}.

\subsection{Morse theory for surfaces without boundary}

Let $S\subset \mathbb{R}^N$ be a smooth compact surface without boundary. The goal is to decompose any given $S$ as in Figure \ref{esfera_morse}. In order to do that, we will compute its Morse-Smale complex, which in the case of a surface only consists of $0$-cells (points), $1$-cells (curves) and $2$-cells (topological disks). 

\smallskip

As explained before, a map $f : S \to \mathbb{R}$ is {\em Morse} if it is $\mathcal C^2$, has only finitely many critical points (i.e. points $p$ where $\text{d}_pf = \nabla f(p) = 0$), and at all of these its Hessian has rank $2$. 

\begin{figure}[h]
	\centering
	\includegraphics[width=0.6\linewidth]{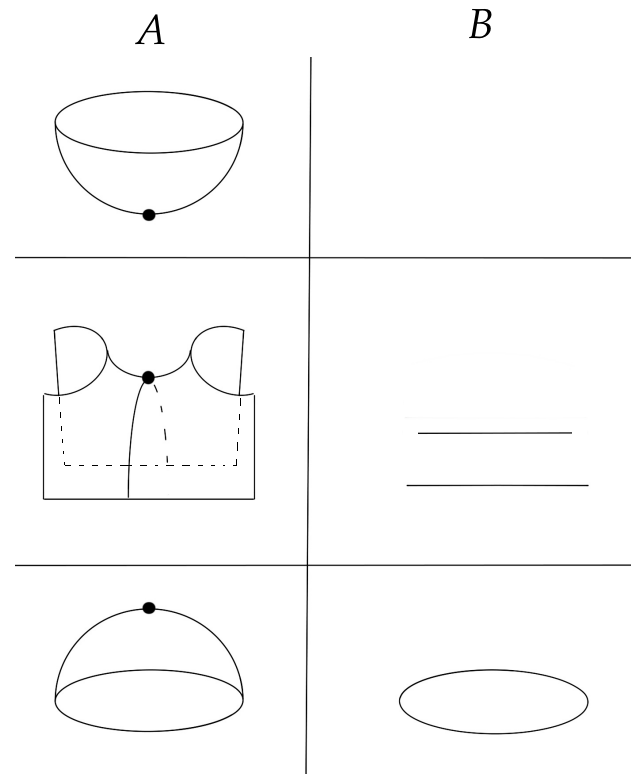}
	\caption{Three types of critical points $p$ (from top to bottom: minima, saddles and maxima) and their Morse data $(A(p),B(p))$ for surfaces without boundary.}
	\label{cirugias_sinborde}
\end{figure}

\begin{definition}[Morse data]\label{morse_data}
	For each critical point $p\in S$, the Morse data are the pair of sets $\left(A(p),B(p)\right)$ where		
	\begin{equation*}
		A(p) := \mathcal{B}(p) \cap f^{-1}\left([f(p)-\epsilon,f(p)+\epsilon]\right),
	\end{equation*}
	with $\mathcal{B}(p)\subset \mathbb{R}^N$ a closed ball around $p$ of sufficiently small radius, the value of $\epsilon > 0$ is such that there are not more critical points of $f$ in $f^{-1}\left([f(p)-\epsilon,f(p)+\epsilon]\right)$ and $ B(p) := \mathcal{B}(p) \cap f^{-1}\left(f(p)-\epsilon\right)$. Notice that $B(p)\subseteq\partial A(p)$. See Figure \ref{cirugias_sinborde} for examples of the sets $A,B$.
\end{definition}
\begin{theorem}[Main theorem of Morse theory]
	Let us now denote $f_{\leq c} = f^{-1}\left((-\infty,c]\right), f_c = f^{-1}(c)$. Then (see [\cite{Hirsch1976DiffTopo}]) as $c\in\mathbb{R}$ increases:
	\begin{description}
		\item[A] If $c_1 < c_2$ and there are no critical values of $f$ between them, then $f_{\leq c_1}$ and $f_{\leq c_2}$ have the same topology (they are actually diffeomorphic).
		
		\smallskip
		
		\item[B] If there is a single critical point $p\in S$ such that $c_1 < f(p) < c_2$ then $f_{\leq c_2}$ is obtained, up to diffeomorphism, from $f_{\leq c_1}$ by attaching the cell $A(p)$ along $B(p)$, i.e. $f_{\leq c_2}\simeq \left(f_{\leq c_1}\cup A(p)\right)\slash\sim$ where the equivalence relation is given by identifying $B(p)\subseteq f_{\leq c_1}$ with points of $\partial A(p)\supseteq B(p)$.  
	\end{description}
	
\end{theorem}

The previous theorem is useful because it allows us to deduce how to couple each type of critical point with the Morse cells that will give us the sought decomposition of the surface under study (e.g. Figure \ref{esfera_morse}). Since $H(f)$ has rank 2 at $p$, we can only have three types of critical points (see Figure \ref{cirugias_sinborde}) according to their index (number of negative eigenvalues of $H(f)$):  
\begin{enumerate}
	\item Minima: $A(p)$ is homeomorphic to a disk and $B(p) = \emptyset$. In this case there is no surgery, the cell $A(p)$ just appears. This cell retracts to a point, the local minimum, which will be a $0$-cell in the Morse-Smale complex.
	\item Saddles: $A(p)$ is a quadrilateral (homeomorphic to a disk) and $B(p)$ consists of two segments. Two opposite sides of $\partial A(p)$ are identified with $B(p)$ (see Figure \ref{cirugias_sinborde}). The attachment of the cell $A(p)$ along $B(p)$ is homotopy-equivalent to the attachment of a $1$-cell, namely the medial axis of $A(p)$ to the middle points of $B(p)$.
	\item Maxima: $A(p) = D$ is homeomorhic to a disk and $B(p) = \partial D$ is its boundary. The attachment map identifies $\partial A(p)$ with $B(p)$. This surgery adds a $2$-cell to the complex.
\end{enumerate}

In summary, minima generate $0$-cells of the complex, saddle points $1$-cells and maxima $2$-cells. For instance, in Figure \ref{esfera_morse} there is one $0$-cell (point $a$), one $1$-cell (the closed curve passing trough $a$ and $b$) and two $2$-cells (one containing $d$ and the other $c$).

\subsection{Morse theory for surfaces with boundary}\label{sec:morse_theory_bnd}

Morse theory also extends to manifolds with boundary via stratified spaces [\cite{Goresky1988StratifiedMT}]. Since stratified Morse theory is less well known and in [\cite{Goresky1988StratifiedMT}] no explicit construction is given for manifolds with boundary, in this short section and in the appendix we give more details on how the boundaries affect the cellular decomposition and the transition between level sets using the terminology presented in the previous section for the case without boundary.

\begin{definition}[Morse function]\label{def:morse_bnd}
	We say that a $\mathcal C^2$ map $f : M \to \mathbb{R}$ defined in a manifold $M$ with boundary $\partial M$ is {\em Morse} if 
	
	\begin{enumerate}
		\item it is Morse in the interior of $M$,
		\item its restriction to $\partial M$, $g := f_{|\partial M}$ is also a Morse function,
		\item if $p\in \partial M$ is a critical point of $g$ then $\ker(\text{d}_p f) = \{0\}$.
	\end{enumerate}
	
\end{definition}


\begin{figure}[h]
	\centering
	\includegraphics[scale=.85]{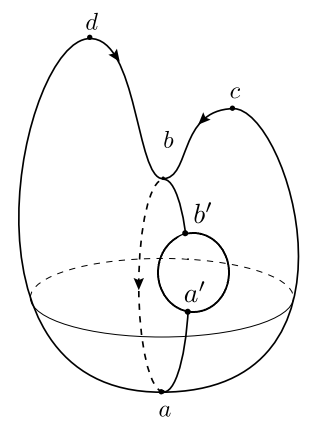}
	\caption{Critical points of the Morse-Smale function $f(x,y,z) = z$ on an example surface with boundary. Notice that now we have a boundary maximum $b'$ and a boundary minimum $a'$ since we have removed a disk. \label{esfera_morse_borde}}
\end{figure}

As in the previous section we will focus in the case $M = S$ is a surface. Notice that now we can have critical points of $f_{|\partial S}$ in the boundary curves $\partial S$ that are not critical points of $f$ in the whole of $S$. Those are only 2 new cases which we call \textit{boundary} minima and maxima (since $\partial S$ is a curve, there can not be saddle points of $g$, see Figure \ref{esfera_morse_borde}). These critical points will be responsible for generating new Morse cells depending on whether they are also local maxima or minima of the full surface or not (see Section \ref{summary_complex}). On a technical side, the third condition of the previous definition rules out the possibility that critical points located at $\partial S$ are saddle points of $f$ in $S$.

\subsection{Construction of the Morse-Smale complex}\label{summary_complex}

We now make a summary of the effect of each critical point on the cell complex once we have made the appropriate deformation retracts. Recall that $0$-cells are points, $1$-cells are curves and $2$-cells are topological disks. Boundary curves of $\partial S$ will be part of the complex as $1$-cells.

\begin{enumerate}
	\item Interior maximum: attach a $2$-cell to the $1$-cells as before.
	\item Interior saddle: attach a $1$-cell to the $0$-cells as before or to a point of the boundary (this point becomes a new $0$-cell).
	\item Interior minimum: add a $0$-cell to the skeleton as before.
	\item Local maximum of $S$ located in $\partial S$: attach a $1$-cell to a $0$-cell (on the boundary) and attach a $2$-cell to the $1$-cells.
	\item Boundary maximum (not of $S$): attach a $1$-cell to a $0$-cell (on the boundary).
	\item Local minimum of $S$ located in $\partial S$: add a $0$-cell to the skeleton (this point will be on the boundary).
	\item Boundary minimum (not of $S$): add a $0$-cell (boundary point) to the skeleton and attach a $1$-cell to a $0$-cell or to a point of the boundary (this point becomes a new $0$-cell).
\end{enumerate}

In order to construct the complex, we first add all the $0$-cells (all interior, boundary minima and possibly some points of the boundary), then the $1$-cells (the boundary curves, the $1$-cells corresponding to each saddle point and the $1$-cells joining boundary minima to local minima or boundary points) and finally add the $2$-cells corresponding to each local maximum.

\remark A delicate point that we have not discussed yet, but will be crucial for our algorithm, is how to determine which cells attach to which cells, { and to find explicitly the attachment maps}. This problem will be addressed in detail for point-clouds in Section \ref{sec_morse_celdas}.

\section{Structure of the point-cloud} \label{sec_local}
Before we can define the flow of a Morse function on a given point-cloud $X$, we will need to give it a local structure. This is done by finding local neighbors to each point, which allows us to estimate tangent planes to $X$ and will be very important later to recognize boundary points.

\subsection{Neighbors identification}\label{sec:vecinos}

\begin{figure}[h]
	\begin{center}
		\includegraphics[scale=0.25]{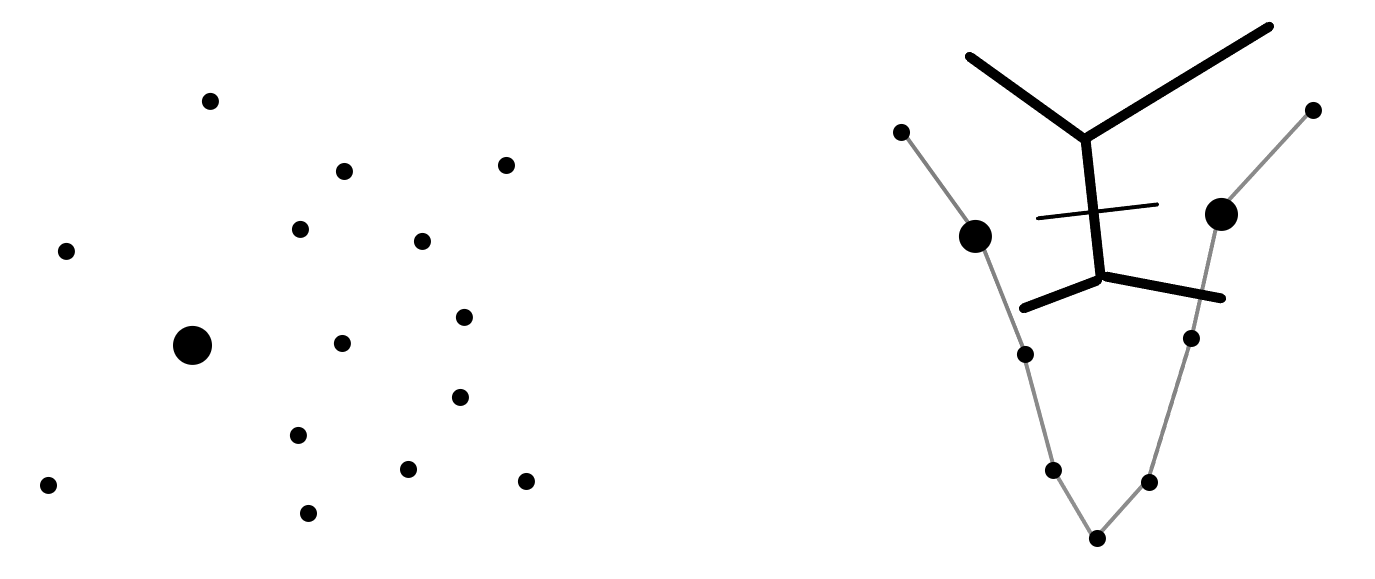}
	\end{center}
	\caption{Typical problems associated to \textit{k}-nearest neighbors (left) and Voronoi neighbors (right). On the left, most closest points to $v$ (in bold) are clustered at one side of it. On the right, vertices that are too far apart from each other belong to neighboring cells. \label{vecinos_nube}}
\end{figure}

The first step is the identification of a set of neighbors of each point $v$ in the cloud $X \subset \mathbb R^N$. There are two classical approaches:

\begin{enumerate}
	\item \textit{k}-nearest neighbors (KNN): given a value for $k$ and a point $v$, the $k$ nearest points $\{v_1,\dots v_k\}$ with respect to the Euclidean distance are declared as its neighbors. This is quite efficient to compute but runs into problems when the point-cloud has irregular densities and $k$ is not big enough, e.g. when all the closest points to $v$ are clustered at one side of it and do not \textit{enclose} the vertex (see Figure \ref{vecinos_nube}, left).
	
	\item Voronoi-Delaunay neighbors: we perform Voronoi's cellular decomposition of the point-cloud and then declare as neighbors of $v$ the points $v_i$ belonging to neighboring cells (i.e. those connected to $v$ by an edge in the Delaunay triangulation of the cloud). This has the virtue of enclosing the vertex $v$ even with irregular densities, but it can be expensive to compute and may produce neighbors too apart from each other (see Figure \ref{vecinos_nube}, right). 
\end{enumerate}

Therefore we merge these two criteria, and declare two points as neighbors when (i) each point is among the $k$-nearest neighbors of the other and (ii) their Voronoi cells in the decomposition of the ambient space $\mathbb R^N$ induced by $X$ are adjoining. In order to be efficient, inspired by \textit{sphere packing theory} [\cite{gensane2004dense}], we first choose a $k = 12$ to compute the $k$-nearest points and then only keep as neighbors the vertices that are connected to $v$ in the Delaunay triangulation of these few points. Finally the relationship of neighborhood is made symmetric by reciprocating neighboring relationships where needed. The neighbors of $v$ will be denoted by $\Neigh(v)$. This produces a  (locally non-planar) graph { of neighbors}, which gives an idea of the local structure of $X$, but which will be in general very complicated.

\subsection{Discrete curvature filter}\label{sec:filtro}

In order to avoid the proliferation of critical points of the Morse-Smale function that in turn would generate a very high number of Morse cells, once we have identified neighbors of each point, we apply a \textit{discrete-curvature filter} { based on the combinatorial Laplacian of the graph of neighbors} to the point-cloud. This means that we substitute each point $v$ for a weighted average of its position and the location of its neighbors:

\begin{equation*}
	h(v) = \alpha v + (1 - \alpha)\frac{1}{|\Neigh(v)|}\sum_{v_i\in\Neigh(v)}v_i.
\end{equation*}
When applied a small number of times this filter defines a bijection between the original cloud and the filtered one, which preserves the topology of the underlying surface (see [\cite{CraneCurvatureFlow}] for a thorough discussion of topology preserving curvature flows applied to triangle meshes). Hence, the decomposition we find for the filtered case will still be valid and topologically accurate for the original cloud. The application of this filter is not always needed, being most relevant when the point-clouds present a lot of noise or a high level of local variability.

\subsection{Tangent space estimation}\label{sec:normales}
This task is performed through \textit{Principal Component Analysis} [\cite{Hoppe1992}]: if the point $v$ and all its neighbors $\Neigh(v) = \{v_1,\dots v_k\}$ were co-planar we would have that for every $i$: $\langle \vec{vv_i},n_j \rangle = 0$, where $n_j$ are all the normal vectors to the surface at $v$ (recall that in general we are in $\mathbb R^N$). Since in general this will not be the case, we find the $n_j$'s by minimizing the function  $\sum_j\sum_{i = 1}^{k}\langle \vec{vv_i},n_j \rangle^2$. This is equivalent to finding the regression plane through $v$ in the least squares sense, and it can be done efficiently by means of a \textit{singular value decomposition} of the matrix with vectors $\vec{vv_i}$ as rows: the right singular vectors corresponding to the 2 largest singular values define the tangent plane, whereas the rest give us the normal directions.


\subsection{Boundary recognition} \label{sec:borde}

\begin{figure}[h]
	\begin{center}
		\includegraphics[width=0.95\linewidth]{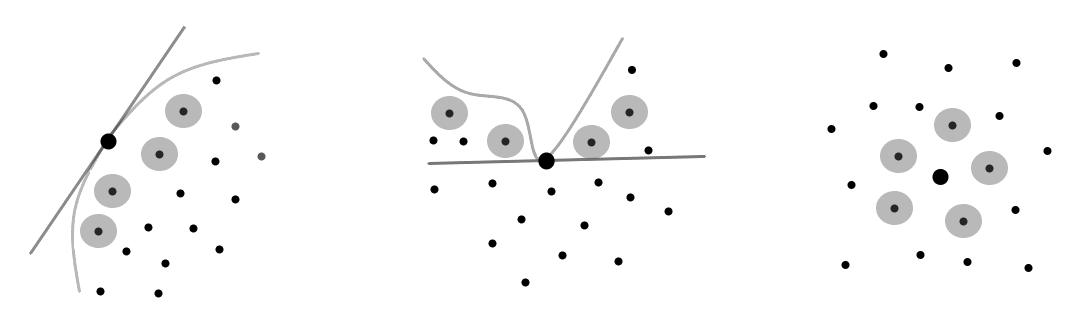}
	\end{center}
	\caption{In principle, a boundary point can be identified easily because after projecting it and its neighbors on the tangent plane, they cluster in a semi-space of $\mathbb{R}^2$ (left panel). Nevertheless, this is not always the case for every boundary point (middle panel). To overcome this, we declare a point as being on the boundary when none of the plane projections of it and its neighbors in all their respective tangent planes enclose the point (right panel).\label{min_max_borde}}
\end{figure}

Once we have an estimation of the tangent spaces, in principle a boundary point of the surface can be easily identified because after orthogonally projecting it and its neighbors on its tangent plane, they cluster in a semi-space (see the first panel of Figure \ref{min_max_borde}). Nevertheless this method is difficult to implement robustly (see the second panel of Figure \ref{min_max_borde}).

\begin{figure}[htb!]
	\begin{center}
		\includegraphics[width=0.6\linewidth]{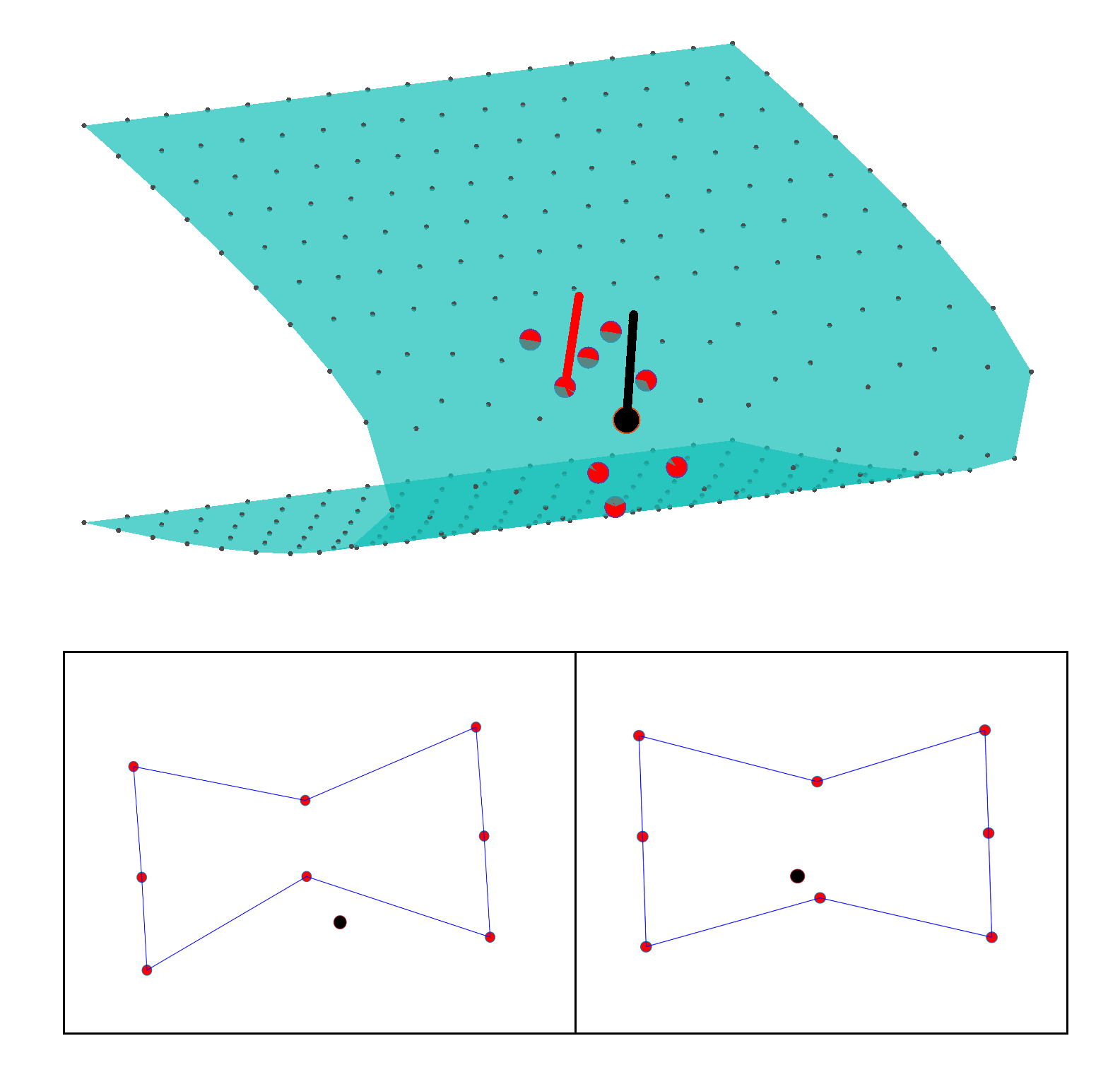}
	\end{center}
	\caption{In interior points with high curvature the projections of the neighbors (red) of a point (black) into the tangent plane of the point may not enclose said point (lower panel, left). In order to avoid false positives we also project the point and its neighbors into the tangent planes estimated for the neighboring points (lower panel right).\label{false_positives}}
\end{figure}

In order to obtain a robust detection, the idea will be to declare points as lying on the boundary only when the projections do not enclose the point. In points with high curvature where the tangent plane may not be perfectly estimated (or equivalently the normal vectors; for a discussion of this phenomenon, see [\cite{NormalEstimation}]) the previous method can give false positives (see Figure \ref{false_positives} lower panel, left). In order to overcome this difficulty, we will also project the point and all its neighbors in the tangent planes estimated for the neighbors. We will build a graph for every projection and only declare $v$ as boundary point when none of the graphs enclose $v$ (see Figure \ref{false_positives} lower panel, right).

Now we proceed to describe our method in detail: let $v\in X$, and $T_v S$ be the estimated tangent plane at $v$. We follow the following steps:
\begin{enumerate}
	\item Given $\{v_1,\dots v_k\}$ the neighbors of $v_0 := v$ we project the $k+1$ points on the planes $T_{v_j} S$ for $j=0,\ldots k$. These projections will be denoted by $\pi_j(v_i)$.
	\smallskip
	\item We create a plane graph $G_j$ with the projected points for every $j=0,\ldots k$, where we add an edge between  $\pi_j(v_i)$ and $\pi_j(v_l)$ only when $v_i,v_l$ are themselves neighbors in the cloud and $i,l\neq 0$.
	\smallskip
	\item We declare the vertex $v_0$ as a boundary point only when none of the plane graphs $G_j$ enclose $\pi_j(v_0)$ for every projection to $T_{v_j} S$.
\end{enumerate}

\remark In point-clouds with irregular densities the previous algorithm can still give false positives because of small boundary holes. In that case clusters of boundary points with small diameter (i.e. clusters contained in small balls) can be discarded.

\subsection{Reconstruction of curves}\label{sec:param_curvas}

Curve reconstruction from a point-cloud sample is used for boundary parametrization, and later for level set reconstruction when we intersect the gradient flow graph
with level hyperplanes. We employ a variant of the NN-Crust algorithm described in [\cite{Dey2006CurveAS}], which can be called KNN-Crust: we limit the search of edges to the $k$-nearest neighbors on the curve, usually with $k = 5$. The rest of the reconstruction follows as in [\cite{Dey2006CurveAS}]. Thus, from now on, we assume that boundary curves of the cloud (if present) have been reconstructed and parametrized.

\section{Morse function and Morse cells}\label{sec_morse_celdas}

Once we have a notion of neighborhoods in the cloud, we are able to define Morse functions, and more importantly their flows. This in turn allows us to compute critical points and from them the Morse cells that form the skeleton of the topological decomposition of the point-cloud. The final step will be to study how these cells attach to each order, in order to recover the full topology of the underlying surface.

\subsection{Morse function and its flows}\label{sec:up_down}
We define $f:X\rightarrow\mathbb{R}$ as the height function $f(v) = \langle v,\nu \rangle$ where $\nu$ is a fixed unitary direction { vector}. We try several $\nu$ randomly until all the $\{f(v)\}_{v\in X}$ are different and the number of local maxima of $f$ (the future number of 2-cells) is small. In the presence of boundaries we also choose the direction $\nu$ that minimizes the number of critical points at the boundaries. This keeps the quantity of boundary $1$-cells small.

\smallskip

Then, following [\cite{Gao2008MorseSmaleD,Zhu2009TopologicalDP}] we define the \textbf{upward flow} of $f$ as the function $\Up:X\rightarrow X$ such that

\begin{equation}
	\Up(v) = \text{argmax}_{v_k\in\Neigh(v),\;f(v_k)>f(v)}\frac{f(v_k) - f(v)}{||v_k - v||}.
\end{equation}

Analogously, the \textbf{downward flow} is the function $\Down:X\rightarrow X$ defined by 

\begin{equation}
	\Down(v) = \text{argmin}_{v_k\in\Neigh(v),\;f(v_k)<f(v)}\frac{f(v_k) - f(v)}{||v_k - v||}.
\end{equation}

When $f(v) > f(v_k)$ for every $k\in\Neigh(v)$, we have a local maximum and write $\Up(v) = v$. Likewise, when $f(v) < f(v_k)$ for every $k\in\Neigh(v)$, we have a local minimum and $\Down(v) = v$.

\smallskip

Since we have identified and parametrized the boundary curves of $X$, we can define the two flows $\partial\text{Down}(\cdot)$ and $\partial\text{Up}(\cdot)$ analogously (considering that each boundary point has two natural neighbors in $\partial X$) so that they send boundary points to boundary points. Then we can easily compute boundary maxima (resp. minima). Notice that in general these points do not need to be local maxima (resp. minima) of the full cloud $X$.

\subsection{Hyperplane sections and computation of critical values}\label{sec:plane_sections}
Next, we perform $n$ level set intersections at equally spaced levels $c_i=c_0 + i \cdot h$, where $c_n = \max f(X)$ and $c_0 = \min f(X)$. { The separation $h$ among levels is the size threshold over which we will be able to detect topological features.} For surfaces with boundaries, however, it may be preferable to choose levels that are not equally spaced, as discussed in Remark \ref{rmk:level}. This process involves intersecting the oriented graph $G_{\text{down}}$, with vertices $X$ and edges defined by $\Down(\cdot)$ (i.e., two vertices $v, w \in X$ share an edge if $w = \Down(v)$), along with the boundary curves, with the hyperplanes $H_{c} = {p \in \mathbb{R}^N: p \cdot \nu - c = 0}$ (see Figure \ref{morse_min_max}). These intersections $\varGamma(c) = G_{\text{down}} \cap H_{c}$ yield plane point-samples of one-dimensional curves (possibly with boundary, i.e., intervals whose endpoints correspond to boundary points of $X$ as identified in Section \ref{sec:param_curvas}), which we can reconstruct and parametrize. By construction, we can trace the correspondence of points between different levels, $\varGamma(c_{i+1})$ and $\varGamma(c_{i})$, using the flow $\Down(\cdot)$. Changes in the number or topological type of the connected components of these level sets indicate that we have crossed critical points of $f$. We will analyze each possible scenario in detail.

\medskip

In order to get the correspondence of points between the levels $\varGamma(c_{i+1})$ and $\varGamma(c_{i})$ using $\Down(\cdot)$, we may need to apply the flow more than once depending on the size $h$ of the jump between level sets that we have defined. The explicit correspondence is obtained as follows: for each node $v$, we consider the trajectory given by the sequence $v^n = \Down(v^{n-1})$ where $v^0 = v$ and then we intersect this polygonal curve with the planes $H_{c_{i+1}}$ and $H_{c_i}$ respectively.

\remark \label{rmk:level} When the surface has non-empty boundary we select level sections that are not necessarily equispaced, so that between two consecutive critical points of $f$ there is always a regular level set (without any critical value) in between. 

\subsubsection*{Case without boundary}
When the underlying surface $S$ of the point-cloud $X$ has no boundary, the reconstructed curves $\varGamma(c)$ are all homeomorphic to the disjoint union of $\mathbb{S}^1$ and hence only 3 changes may happen:

\begin{enumerate}
	\item \textit{Passing a maximum}: when going down from $\varGamma(c_{i+1})$ to $\varGamma(c_{i})$ a new connected component appears (see Figure \ref{morse_min_max}, left). These new points can \textbf{not} be reached from points of $\varGamma(c_{i+1})$ using the flow $\Down(\cdot)$.
	
	\smallskip
	
	\item \textit{Passing a minimum}: when going down from $\varGamma(c_{i+1})$ to $\varGamma(c_{i})$ an existing connected component disappears (see Figure \ref{morse_min_max}, right). These points do \textbf{not} go to points of $\varGamma(c_{i})$ using the flow $\Down(\cdot)$.
	\begin{figure}[H]
		\begin{center}
			\includegraphics[scale=0.2]{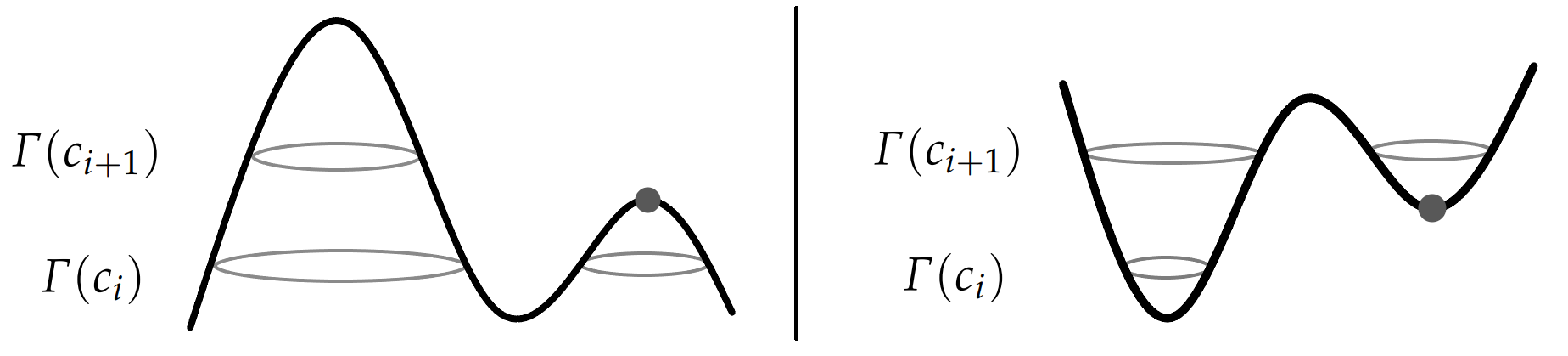}
		\end{center}
		\caption{\label{morse_min_max}\textit{Left}: Passing a maximum point when following the downwards flow: a connected component appears. \textit{Right}: Passing a minimum when following the downwards flow: an existing connected component disappears.}
	\end{figure}
	
	\item \textit{Passing a saddle}: when going down from $\varGamma(c_{i+1})$ to $\varGamma(c_{i})$ a connected component appears or disappears (see Figure \ref{morse_silla}). Points on any component of $\varGamma(c_{i+1})$ have images in every component of $\varGamma(c_{i})$ using the flow $\Down(\cdot)$.
	
	\begin{figure}[H]
		\begin{center}
			\includegraphics[scale=0.2]{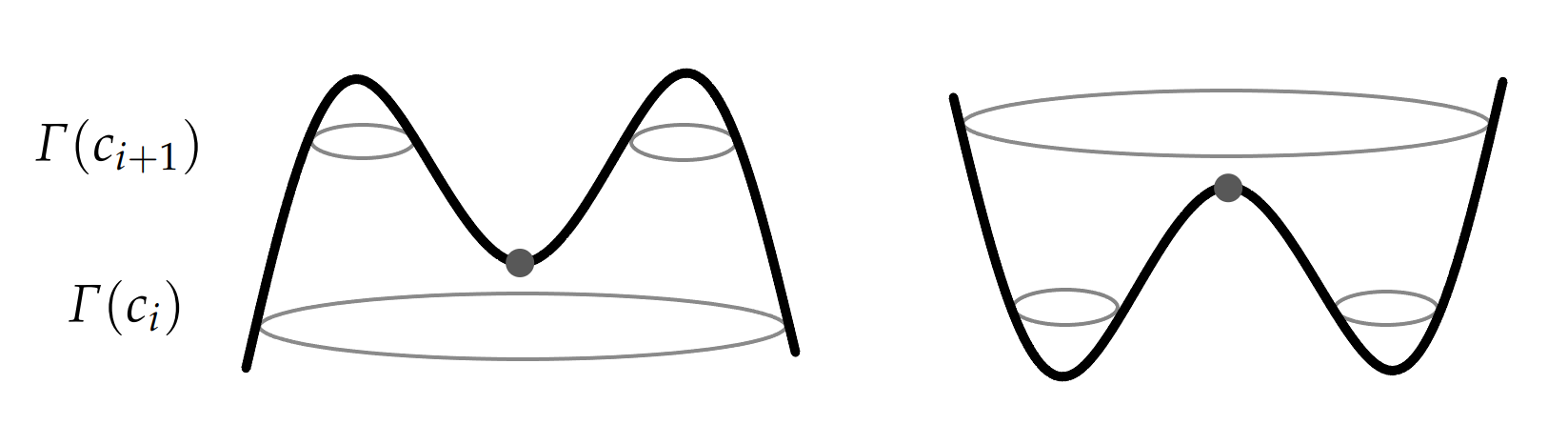}
		\end{center}
		\caption{\label{morse_silla}Passing a saddle point when following the downwards flow: a connected component appears or disappears.}
	\end{figure}
	
\end{enumerate} 

\subsubsection*{Case with boundary}
When the underlying surface $S$ of the point-cloud $X$ has a boundary $\partial S$, the reconstructed curves $\varGamma(c)$ are homeomorphic to a union of $\mathbb{S}^1$ and closed intervals $[0, 1]$. These intervals always join two points of $\partial S$, introducing a bordism equivalence relation in $\partial S\cap H_c$. Therefore, we have the three previous cases, plus two new ones, since now we can have minima and maxima at the boundaries. Saddle points can not be located at the boundaries since that would violate the third condition of Definition \ref{def:morse_bnd} as already explained. We will make a distinction between (local) minima (resp. maxima) of the whole point-cloud, and boundary minima (resp. maxima) which are points that are critical when we restrict $f$ to the boundary curves, but not in the whole surface. 

\smallskip

We can have the following 5 cases:

\begin{enumerate}
	\item \textit{Passing a local maximum}: this is similar as before, when going down from $\varGamma(c_{i+1})$ to $\varGamma(c_{i})$ a new connected component $\mathbb{S}^1$ or $[0,1]$ whose points can \textbf{not} be reached from $\varGamma(c_{i+1})$ using $\Down(\cdot)$ appears.
	
	\item \textit{Passing a local minimum}: similarly as before, when going down from $\varGamma(c_{i+1})$ to $\varGamma(c_{i})$ an existing connected component $\mathbb{S}^1$ or $[0,1]$ whose points do \textbf{not} go to $\varGamma(c_{i})$ using $\Down(\cdot)$ disappears.
	
	\item \textit{Passing a boundary maximum}: when going down from $\varGamma(c_{i+1})$ to $\varGamma(c_{i})$ either a $[0, 1]$ opens in two intervals or a $\mathbb{S}^1$ splits into a $[0, 1]$.
	
	\item \textit{Passing a boundary minimum}: when going down from $\varGamma(c_{i+1})$ to $\varGamma(c_{i})$ either a $[0, 1]$ closes to be a $\mathbb{S}^1$ or two $[0,1]$ merge into one interval.
	
	\item \textit{Passing a saddle}: this is the most complex case, when going down from $\varGamma(c_{i+1})$ to $\varGamma(c_{i})$ a connected component $\mathbb{S}^1$ or $[0, 1]$ appears or disappears. We can also have a case where the number of connected components stays the same but in passing from $\varGamma(c_{i+1})$ to $\varGamma(c_{i})$ the bordism pairing of boundary points of $\partial S$ given by the $[0,1]$ component changes. Points on any component of $\varGamma(c_{i+1})$ have images in every component of $\varGamma(c_{i})$ using the flow $\Down(\cdot)$.
\end{enumerate} 

\begin{figure}[htb!]
	\centering
	\includegraphics[width=0.75\linewidth]{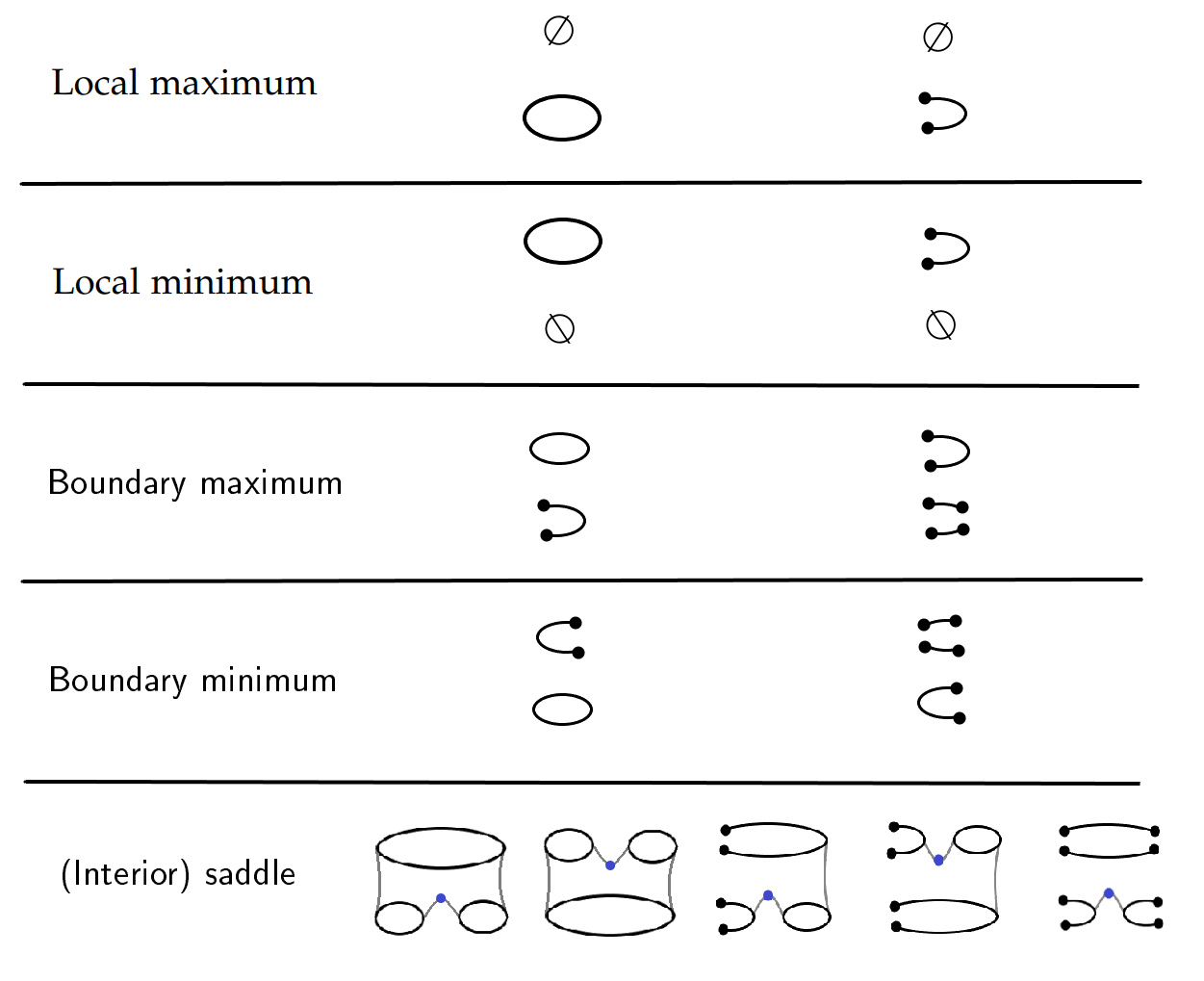}
	\caption{\small Different local level set transformations: the reconstructed curves $\varGamma(c)$ are homeomorphic to a union of  $\;\mathbb{S}^1$ and closed intervals $[0, 1]$.  \label{casos_borde}}
\end{figure}

In Figure \ref{casos_borde} we can see a summary of the generic (i.e. non-degenerate) level set transformations when the level crosses that of a particular type of critical point. 

\subsection{Identification of Morse cells}\label{sec:morse_cells}
In this section we explain how to identify all the \textit{discrete} Morse cells of the complex that give us a cellular decomposition of the point-cloud.

\subsubsection*{$0$-cells}

They correspond to local minima or to boundary minima of $f$, we already found them; they are the fixed points of $\Down(\cdot)$ and $\partial\Down(\cdot)$ respectively. 

\subsubsection*{$1$-cells}

There are three cases that generate the $1$-cells of the skeleton:

\begin{enumerate}
	\item Boundaries: when $S$ has non-empty boundary, the boundary curves are also part of the $1$-skeleton. We already parametrized those. Each boundary maximum gives rise to a $1$-cell that attaches to a $0$-cell (point) located at the boundaries.
	
	\item Saddle points: there are one-dimensional curves that go from one local minimum $m_1$ or boundary point to another (not necessarily distinct) local minimum $m_2$ or boundary point passing through the saddle point. 
	
	\item Boundary minima: there are $1$-cells in the complex not associated to a saddle point or to boundary curves. In this last case they always connect a boundary minimum to a local minimum or boundary point of the cloud. In order to compute this $1$-cell we simply flow down every boundary minimum using $\Down(\cdot)$. 
\end{enumerate}

\remark \label{rmk:0celdas}When $1$-cells introduced by saddle points or boundary minima end at points of the boundary $\partial S$, { or of preexisting $1$-cells, which are not the ends of its original $1$-cell and thus not part of the $0$-skeleton, the original $1$-cell is split into 2 by the meeting point, which is labeled as a new $0$-cell.}

\subsubsection*{Computation of saddles and their $1$-cells}

\begin{figure}[H]
	\begin{center}
		\includegraphics[width=\linewidth]{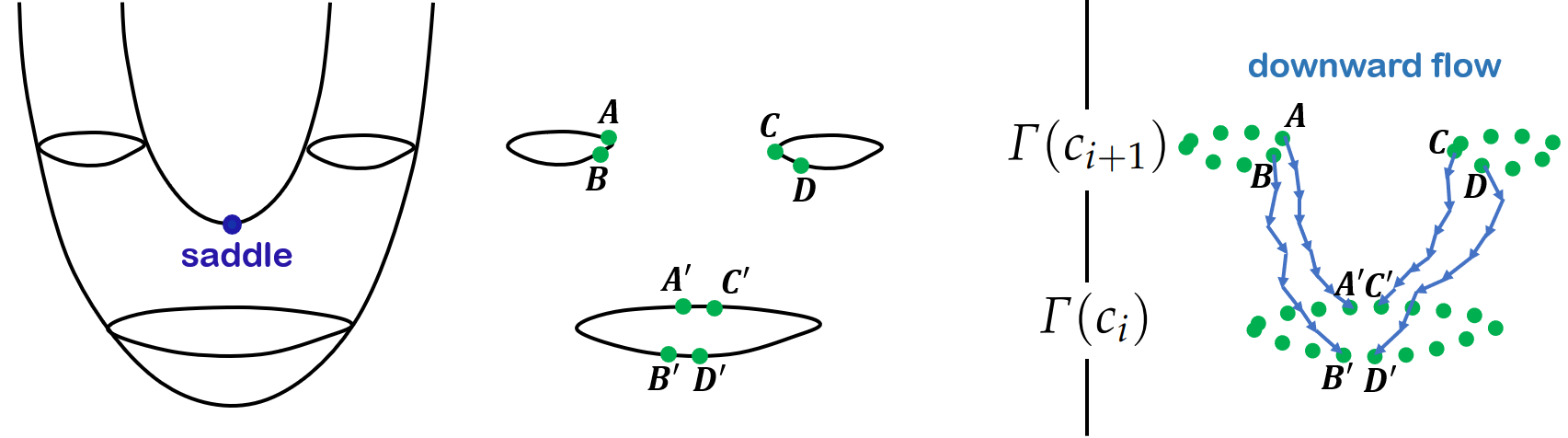}
	\end{center}
	\caption{Change in level set topology when crossing a saddle point in a surface (left) and point-cloud (right): note the change in neighbors among the 4 marked points after applying the flow $\Down(\cdot)$. \label{f:silla}}
\end{figure}

We now explain how to compute saddle points and afterwards their associated $1$-cells. Recall that we know when we are in the presence of a saddle: when going from $\varGamma(c_{i+1})$ to $\varGamma(c_{i})$ a connected component appears, disappears or there is a change of bordism pairing of boundary points. In any case, all points of $\varGamma(c_{i+1})$ have images in $\varGamma(c_{i})$ using the flow $\Down(\cdot)$. Therefore, there exist four points $A,B,C,D$ in $\varGamma(c_{i+1})$ that can be paired by proximity (i.e. are neighbors in the level set curves) as $\{A,B\}$ and $\{C,D\}$ whose images by $\Down(\cdot)$ are now paired differently as $\{A',C'\}$ and $\{B',D'\}$ (see Figure \ref{f:silla}). The saddle point is approximated by the average of the 8 involved points.

One branch of the $1$-cell is obtained by taking the average of the two orbits generated by $\Down(\cdot)$ starting from $\{A',C'\}$. The other branch is approximated in the same manner but starting from $\{B',D'\}$. This averaging operation generates new points not previously in the cloud. All these new points are added to the cloud, declaring as neighbors of $s$ the 8 points used for its computation, and for each new point of the branches its two neighbors in the $1$-cell plus the two points that were used for its computation. The flows $\Down(\cdot)$ and $\Up(\cdot)$ are simply defined at those points by following the newly created trajectories down or up (except the saddle points which are fixed points of both flows). This ensures that this newly defined curve (the $1$-cell) is invariant by both flows.

\subsubsection*{$2$-cells}

There is one $2$-cell for each local maximum of $f$ (recall that boundary maxima only generate $1$-cells, i.e. the boundary curves), we already found them (they are the fixed points of $\Up(\cdot)$), but we will now explain how to deduce which points of the cloud correspond to each $2$-cell (or each maximum). The idea is simply to flow up every point $v\in X$ by $\Up(\cdot)$ and see at which maximum it ends up. In symbols this means that we consider the limit of the sequence $v^n = \Up(v^{n-1})$ where $v^0 = v$ when $n\rightarrow+\infty$ (this limit exists because we have a finite number of points and maxima are fixed by $\Up(\cdot)$). It can happen that the sequence $\{v^n\}_{n=1}^{+\infty}$ converges to a saddle $s$. This can only be allowed for points of the $1$-cells, but must be corrected for the rest: when a point $v$ ends at a saddle $s$ but it is not part of the $1$-cell, we look at the neighbors $\Neigh(v)$ and see in which maximum they finished. The most common maximum among the neighbors is selected as the corrected destination of $v$. It may be necessary to iterate this procedure a finite number of times.

\subsection{Attachment maps of the Morse cells}\label{sec:corona_flow}

Now that we have the skeleton of $S$, i.e. the $0,1,2$-cells, we encounter a delicate problem: figuring out the attachment maps between the cells. We already know how the $1$-cells attach to the $0$-cells. We now discuss how $2$-cells attach to $1$-cells. We must figure out which $1$-cells are the boundary of the $2$-cells, in which order, orientation and how many times they appear: once (when they are part of the boundary of $S$ or when they attach to a different $2$-cell) or twice (when they will be, formally, two different sides of the $2$-cell that are identified). 

\subsection*{Smooth case}

We first describe the process for the smooth case (without boundary) and then explain how to adapt it to point-clouds. Notice that the $2$-cell corresponding to each maximum $\hat{m}$, let us call it $D_{\hat{m}}$, is the set of points that converge to $\hat{m}$ under the flow $+\nabla f$ (in Dynamical System terminology, its stable manifold). The main idea to deduce how the boundary of $D_{\hat{m}}$ attaches to the $1$-cells is to choose a simple closed curve around each maximum and flow it down by $-\nabla f$ until it reaches the $1$-cells. To achieve this properly, we must perturb $-\nabla f$ so that in a neighborhood of the $1$-cells the gradient flow is \textit{transversal} (i.e. not parallel) to the $1$-cells. This is needed since otherwise the points of $D_{\hat{m}}$ would converge to minima under the downwards flow $-\nabla f$. In order to obtain this perturbed $-\nabla\tilde{f}$ we consider an arbitrary small tubular neighborhood around the $1$-cells and a normal vector field to the $1$-cells. These two flows are joined smoothly in the tubular neighborhood with the aid of a partition of unity (i.e. bump functions, see [\cite{Hirsch1976DiffTopo}]).  

\subsection*{Discrete case}

In the discrete case we choose a simple (i.e. without self-intersections) closed curve of neighbors around each maximum (a \textit{crown}) and flow it by $\Down(\cdot)$ until it reaches the $1$-cells. We do this in a way such that when a point of the curve has neighbors that are in the $1$-cell, we stop following the flow $\Down(\cdot)$ and match the point in question to the point in the $1$-cell which results in the most negative downwards slope. This method would work perfectly if simple closed curves were preserved by $\Down(\cdot)$; since this is not the case, we must \textit{repair} the curve every time we flow it down. There are three main situations:

\begin{enumerate}
	\item \textit{Splitting of points}: it may happen that two points $w_1$ and $w_2$ were neighbors but $\Down(w_1)$ and $\Down(w_2)$ are not. In that case we define the sub-graph of neighbors given by points 
	\begin{equation*}
		\{v\in X:\; f(v) \leq \max\left[f(\Down(w_1)),f(\Down(w_2))\right]\},
	\end{equation*}
	and compute the shortest path joining $\Down(w_1)$ to $\Down(w_2)$ (taking as distance of a path that given by the edges of the graph).
	\item \textit{Collapse of points}: it may happen that two points $w_1$ and $w_2$ have the same image $\Down(w_1)=\Down(w_2)$. In that case we simply delete one of the repeated points from the curve.   	
	\item \textit{Creation of spikes}: it may happen that one of two neighboring points $w_1$ and $w_2$ has image $\Down(w_2)=w_1$. This could happen with more than one point at the time. In that case we simply delete the \textit{spiky} points (i.e. $\Down(w_2) = w_1$) in the new curve.
\end{enumerate}

The final problem we face is that the flowing crown may arrive at a stage where it is stationary by the downward flow but some points of it have not reached the $1$-cells. In that case we pair each unmatched point of the crown with the closest (as measured by the edges of the neighboring relationship of the cloud) point of the $1$-cells. 

\smallskip

We now have a matching between the initial crown and the $1$-cells (although not a bijection). But this is enough to deduce which $1$-cells are the boundary of $D_{\hat{m}}$ (only the $1$-cells that are reached), in which order (this can be deduced since the crown is parametrized), the orientation (again thanks to the parametrization) and how many times they appear (once or twice, depending on the matching). This information completely determines the boundary operator of the Morse-Smale complex and hence the homology of the surface (see [\cite{Hirsch1976DiffTopo}]).

\subsubsection*{Attachment maps for surfaces with boundaries}

When the surface $S$ has a boundary $\partial S$, the boundaries of the $2$-cells are no longer only the $1$-cells derived from the saddle points, but also possibly curves of $\partial S$. We distinguish several cases depending on the type of maximum we have:

\begin{enumerate}
	\item Interior maxima (not belonging to $\partial S$): as before we flow down a crown around the maximum until it reaches a point that is either a neighbor of a boundary point or of a $1$-cell. In this way we obtain a matching between the boundary of the $2$-cell and the $1$-cells and boundary curves of $S$.
	\item Boundary maxima (but not a local maximum of $S$): they do not intervene in the attachment maps of $2$-cells to $1$-cells (although they give rise to a $1$-cell in the boundary). 
	\item Local maxima (located at $\partial S$): we consider a \textit{semi-crown} around the maximum $\hat{m}$, meaning that we take an arc of the boundary centered at $\hat{m}$ and complete it with neighboring interior nodes so that we obtain a closed simple curve. Then we flow down this semi-crown in such a way that the arcs of the boundary stay in the boundary (we use the restricted flow $\partial\Down(\cdot)$) and the interior nodes of the curve flow down normally by $\Down(\cdot)$. As before we flow down the semi-crown until each point of it reaches a node that is either a neighbor of a boundary point or of a $1$-cell.
	
\end{enumerate}

For an example of this process in action, see the Supplementary Video 1.

\subsection{Algorithm for the topological decomposition of point-clouds}

To finish this section, we describe our full algorithm in pseudo-code (see Algorithm \ref{algo_morse}), referencing the relevant sections needed to carry out the necessary calculations. The only inputs we need are: the point-cloud $X$, the number of neighbors $k$ to consider and a unitary direction $\nu$. As the final result we get a decomposition consisting of $0$-cells $\mathcal{C}_0$ (points), $1$-cells $\mathcal{C}_1$ (polygonal curves) and $2$-cells $\mathcal{C}_2$ (collection of points); together with their attachment maps $g_1: \mathcal{C}_1\rightarrow \mathcal{C}_0$, $g_2: \mathcal{C}_2\rightarrow \mathcal{C}_1$ telling us which cells are the boundary of other cells.

\begin{algorithm}[H]
	\begin{algorithmic}[1]
		\Require{$X,k,\nu$}
		\State {$\text{Neigh} \gets \text{neighbors}(X,k)$}
		\Comment(Section \ref{sec:vecinos})
		\State {$X \gets \text{curvatureFilter}(X)$}
		\Comment(Section \ref{sec:filtro})
		\State {$\text{Normals} \gets \text{tangentPlanes}(X,\text{Neigh})$}
		\Comment(Section \ref{sec:normales})
		\State {$\partial X \gets \text{boundaryPoints}(X,\text{Neigh},\text{Normals})$}
		\Comment(Section \ref{sec:borde})	
		\State {$\partial S \gets \text{parametrizeCurves}(\partial X)$}
		\Comment(Section \ref{sec:param_curvas})
		\State {$\text{Up}, \text{Down} \gets \text{MorseFlows}(X,\text{Neigh},\nu)$}
		\Comment(Section \ref{sec:up_down})
		\State {$\mathcal{P} \gets \text{criticalPoints}(X,\partial S, \text{Down},\nu)$}
		\Comment(Section \ref{sec:plane_sections})
		\State {$\mathcal{C}_0,\mathcal{C}_1,\mathcal{C}_2 \gets \text{MorseCells}(X,\mathcal{P},\text{Down},\text{Up})$}
		\Comment(Section \ref{sec:morse_cells})
		\State {$g_1,g_2 \gets \text{attachmentMaps}(X,\mathcal{C}_0,\mathcal{C}_1,\mathcal{C}_2,\text{Down})$}
		\Comment(Section \ref{sec:corona_flow})\\
		\Return{$\mathcal{C}_0,\mathcal{C}_1,\mathcal{C}_2,g_1,g_2$}
		\caption{Reconstruction of point-cloud surfaces}\label{algo_morse}
	\end{algorithmic}
\end{algorithm}

\section{Parametrization of the $2$-cells}\label{sec:param2cells}

Once we have each Morse cell and their attachment maps identified, the last problem we face is how to parametrize the $2$-cells, i.e. finding a flat domain $D\subset\mathbb{R}^2$ and a map $\phi:D\rightarrow X$, such that each $\phi(D)\subset X$ corresponds to one of the $2$-cells found before. We will further require $D$ to be a convex polygon and that $\partial D$ is isometric to $\phi(\partial D)$. Once we have a parametrization of each $2$-cell, since they attach well (their boundaries are the $1$-cells), we have a full piece-wise parametrization of $X$. In order to find $D$ and $\phi$, we follow two steps: 

\begin{enumerate}
	\item Take any convex polygon in the plane whose boundary $\partial D$ is isometric to the boundary of the $2$-cell (e.g. a rectangle).
	
	\smallskip
	
	\item Obtain a correspondence of interior points of $D$ mapping to the cloud points in the $2$-cell: for each interior point $v$ in the $2$-cell we want to find an interior point $x = \phi^{-1}(v)$ in $D$. We assume that each point $p_i$ of $D$ (including $\partial D$) is a barycentric combination of its neighbors { (as given by the neighboring relationship of the cloud) with Tutte's weights (all coefficients are equal to $\frac{1}{|\Neigh(v)|}$) or weights inverse to the distance, as proposed by Floater (see [\cite{Floater2002ParameterizationOT}]). This results in a sparse linear system, whose solution gives us the coordinates of interior points of $D$ mapping to the cloud.} 
\end{enumerate}

Once we have these points, we can extend the parametrization to the whole interior of $D$ by taking its Delaunay triangulation with the newly found vertices, and interpolating linearly for the images. Then we can re-mesh (or even quadrangulate) the polygon $D$ if desired. 

\remark\label{r:param} We may wish $\partial D$ to also have the same number of sides of $\phi(\partial D)$ (i.e. the different $1$-cells), their length and their order (these are known in advance and will be determined by the attachment map of the $2$-cell to the $1$-cells). In that case denoting the sides' lengths by $l_1,\dots,l_d$; we can obtain a convex polygon $D\subset\mathbb{R}^2$ whose sides are $l_1,\dots,l_d$ by finding the polygon with maximal area and predetermined sides $l_1,\dots,l_d$ inscribed in a circumference. To do so, an optimization problem is solved in order to find the circumference in question. By a result of Bramagupta [\cite{maley2005areas}], this is done by finding the radius that makes the polygon's interior angles add up to $2\pi$. Once we have $D$, we obtain a bijection between $\partial D$ and the points of the corresponding $1$-cells in $X$. 

\section{Results}\label{sec_results}

In this section we reconstruct five different surfaces, one without boundary and four with it, which pose various challenges to the presented algorithm.  Three of the point-clouds are synthetic whereas the two are real 3D scans, one of an actual textile: a pair of pants. All cases are challenging and interesting for different reasons: the \textit{torus} is really slim and its embedding describes a (2,3)-toric knot which causes far away parts the surface (as measured by geodesic distance) to be really near each other in Euclidean space. The \textit{vest} was obtained by cutting out parts of an ellipsoid in order to obtain a surface with the same topology as an open vest. Therefore, it has positive Gaussian curvature everywhere and very large boundary curves. The \textit{turbine} challenges our algorithm since it presents a non-uniform density distribution of sample points. Furthermore, the \textit{rosette} is a sample cloud from a algebraic surface (the zeros of a polynomial). It has two distinct connected components, very large and at the same time very tiny boundary curves and moreover nontrivial curvature everywhere which causes the appearance of three saddle points. And finally the \textit{pants} are a 3D-scan of a real pair of jeans and thus its point-cloud presents wrinkles, noise and a non-uniform distribution of points. 

\begin{table}[hbt!]
	\centering
	\begin{tabularx}{0.9\textwidth}{Xccccc}
		
		\toprule
		Surface  & $|\partial S|$ & $0$-cells & $1$-cells & $2$-cells & $\mathcal{X}(S)$ \\
		\midrule
		Knotted torus  & 0 & 2 & 4 & 2 & 0\\
		Ellipsoidal vest  & 3 & 5 & 8 & 2 &-1 \\
		Turbine fan blade  & 4 & 7 & 10 & 1 &-2 \\
		Algebraic rosette  & 4 & 12 & 16 & 4 & -1 / +1*\\
		Scanned pants & 3 & 4 & 6 & 1 &-1 \\
		\bottomrule
	\end{tabularx}
    \caption{Reconstructed point-cloud samples and their computed topologies.}\label{tab:topology}
    
	\begin{tablenotes}
		\item \textit{Note}: $|\partial S|$ denotes the number of connected components of the boundary and $\mathcal{X}(S)$ the Euler characteristic computed as $\#0$-cells - $\#1$-cells + $\#2$-cells. 
		
		\item *In the case of the Rosette  $\mathcal{X}(S)$ is computed separately for each connected component.
	\end{tablenotes}
\end{table}

A summary of the results of applying our algorithm to the five point-clouds can be seen in Table \ref{tab:topology}. There we display the number of reconstructed curves of the boundary and the Morse cells found. This information is enough to characterize topologically surfaces with boundary, since with it we can compute the Euler characteristic of the surface [\cite{Hirsch1976DiffTopo,Munkres1984ElementsOA}]. Notice that even though the main connected component of the rosette, the vest and the pants are topologically equivalent (i.e. homeomorphic), their cellular decompositions are completely different because of how they are embedded in $\mathbb{R}^3$ and their different geometries.

\begin{figure}[htb!]
	\centering
	\includegraphics[scale=0.275]{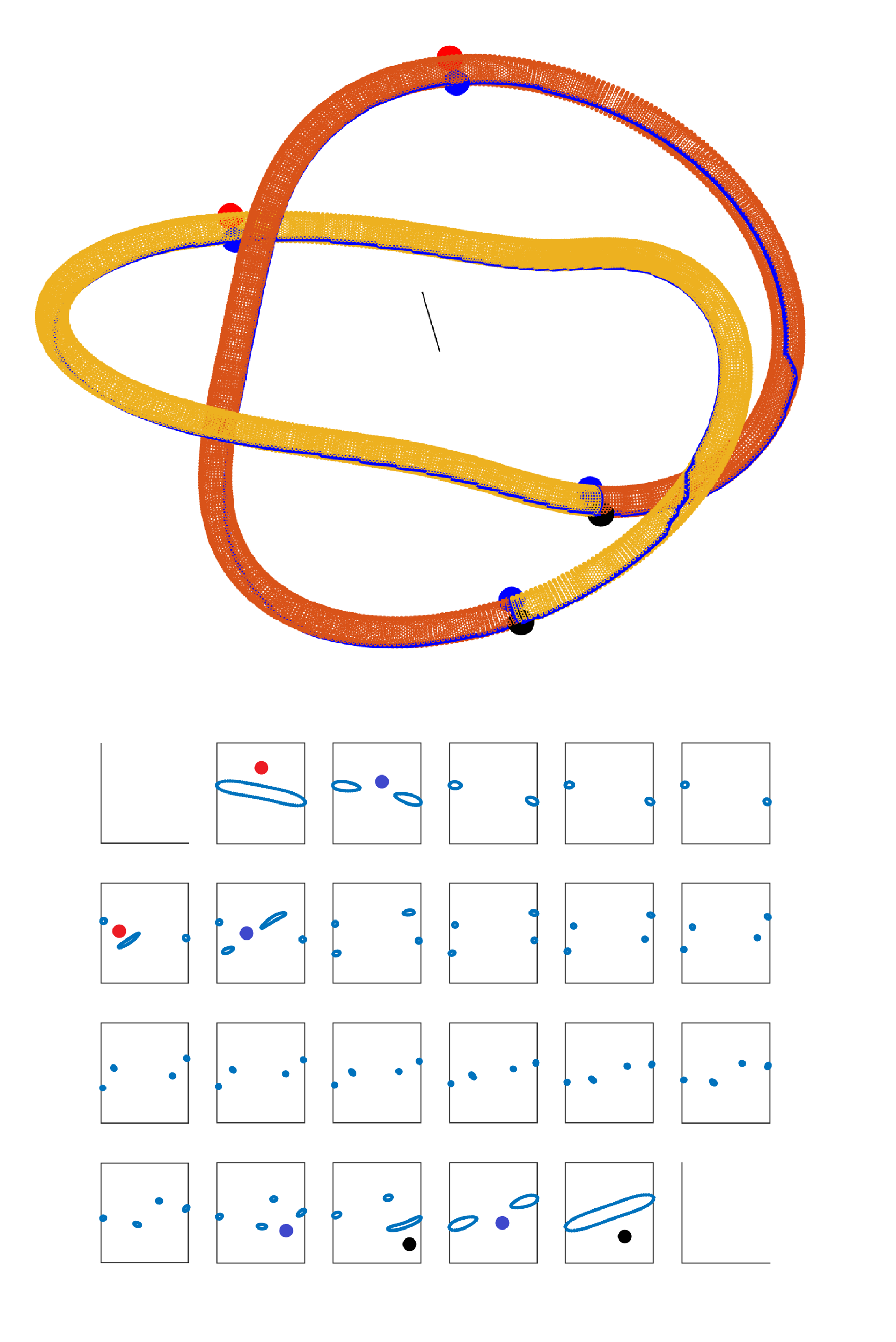}
	\caption{A sampled knotted torus: the black line is the direction of the height function; local maxima, resp. minima, are painted red, resp. black; saddle points are painted blue; their $1$-cells are outlined in blue. On the bottom we plot the level set curves highlighting when critical points appear.} \label{f:nustoric}
\end{figure}

\subsection*{Knotted torus}

Figure \ref{f:nustoric} shows our algorithm applied to a point-sample from a torus embedded in $\mathbb R^3$ along a (2,3)-toric knot. The algorithm correctly detects 2 local maxima, 2 local minima and 4 saddle points for the height function depicted in the figure. Out of a point cloud of 30 000 points, a decomposition of the surface into 8 Morse cells is found (two $0$-cells: the minima, four $1$-cells associated to the saddle points and two $2$-cells, one for each maximum). On the bottom of Figure \ref{f:nustoric} we display the level set curves $\varGamma(c) = G_{\text{down}}\cap H_{c}$ (see Section \ref{sec:plane_sections}). Notice that for this point-cloud we have all possible local transformations of level-set curves for surfaces without boundary (see Figure \ref{casos_borde}).

\begin{figure}[htb!]
	\centering
	\includegraphics[scale=0.325]{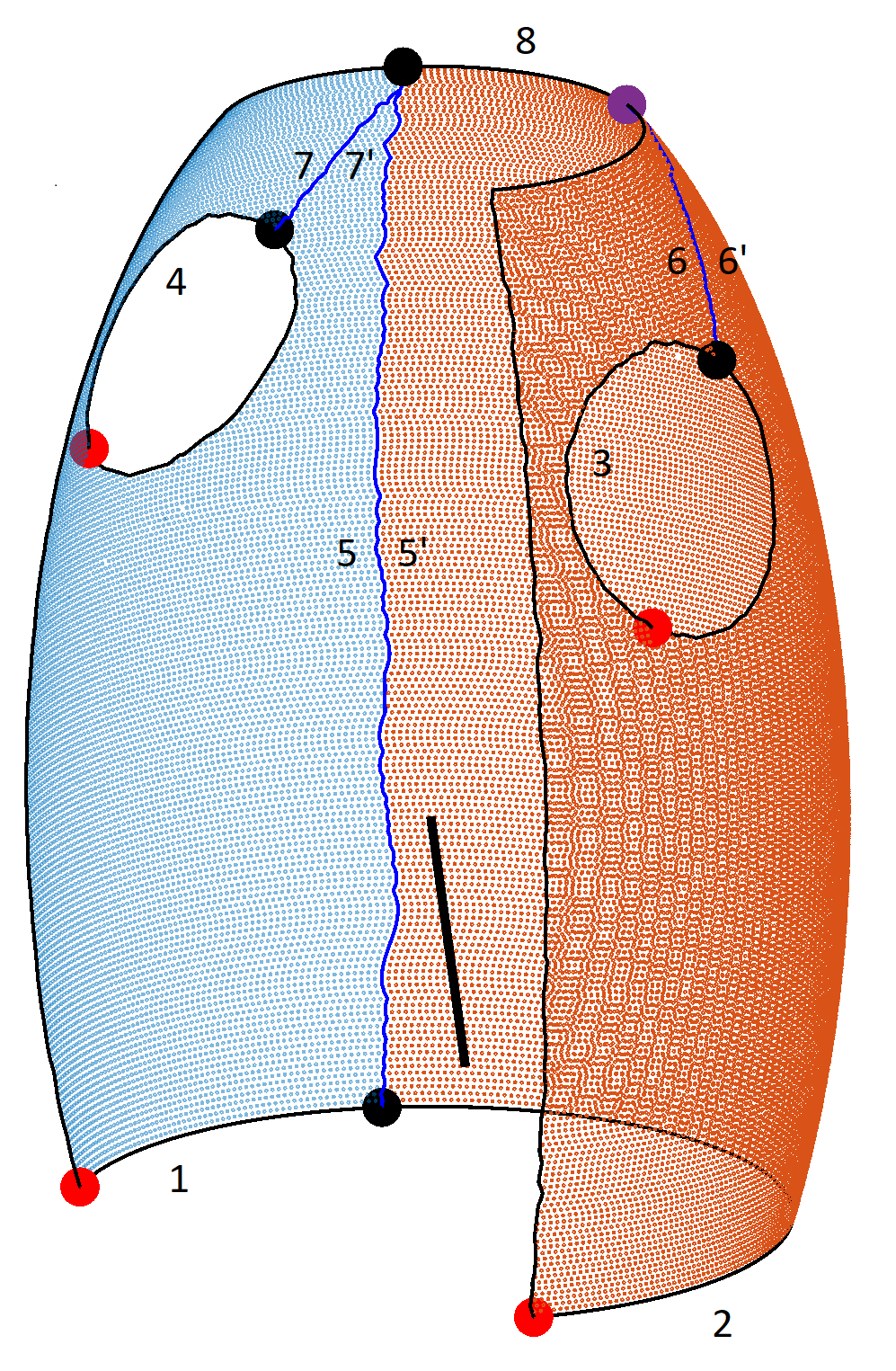}
	\caption{A sampled vest: the black line is the direction (downwards in this case) of the height function; maxima are painted in red, minima in black; $1$-cells corresponding to boundary minima are outlined in blue and the boundary curves in black. One purple point where a $1$-cell meets a boundary curve is added to the decomposition. The numbers correspond to the different formal $1$-cells that, when identified (e.g. 7 with 7'), reconstruct the entire surface from 2 pieces homeomorphic to disks.} \label{f:armilla}
\end{figure}

\subsection*{Ellipsoidal vest}

Figure \ref{f:armilla} shows our algorithm applied to a sample of 36 000 points from an embedded vest in $\mathbb{R}^3$. The point-cloud was generated by cutting sections of an ellipsoid to create a surface with the same topology as an open vest. After successfully detecting and parametrizing the boundary, the algorithm correctly identifies 2 local maxima, 2 boundary maxima, 1 local minimum, and 3 boundary minima for the height function (pointing downwards) shown in the figure. All critical points are located at the boundary. Additionally, one new point (shown in purple) is introduced where a $1$-cell intersects a boundary curve (see Remark \ref{rmk:0celdas}).

Then, a decomposition of the surface into 17 Morse cells is found (five $0$-cells, eight $1$-cells and two $2$-cells). In order to deduce how the two $2$-cells attach and which $1$-cells are their boundary, we apply the curve flow explained in Section \ref{sec:corona_flow}. This process in action for one of the $2$-cells can be visualized in the Supplementary Video 1. Thus, we deduce how the cells attach with each other (e.g. the $1$-cell number 5 appears on both $2$-cells and it is precisely one of the curves where they attach, see Figure \ref{f:armilla}). From this we recover the entire topology of the vest.

\begin{figure}[H]
	\centering
	\includegraphics[scale=0.75]{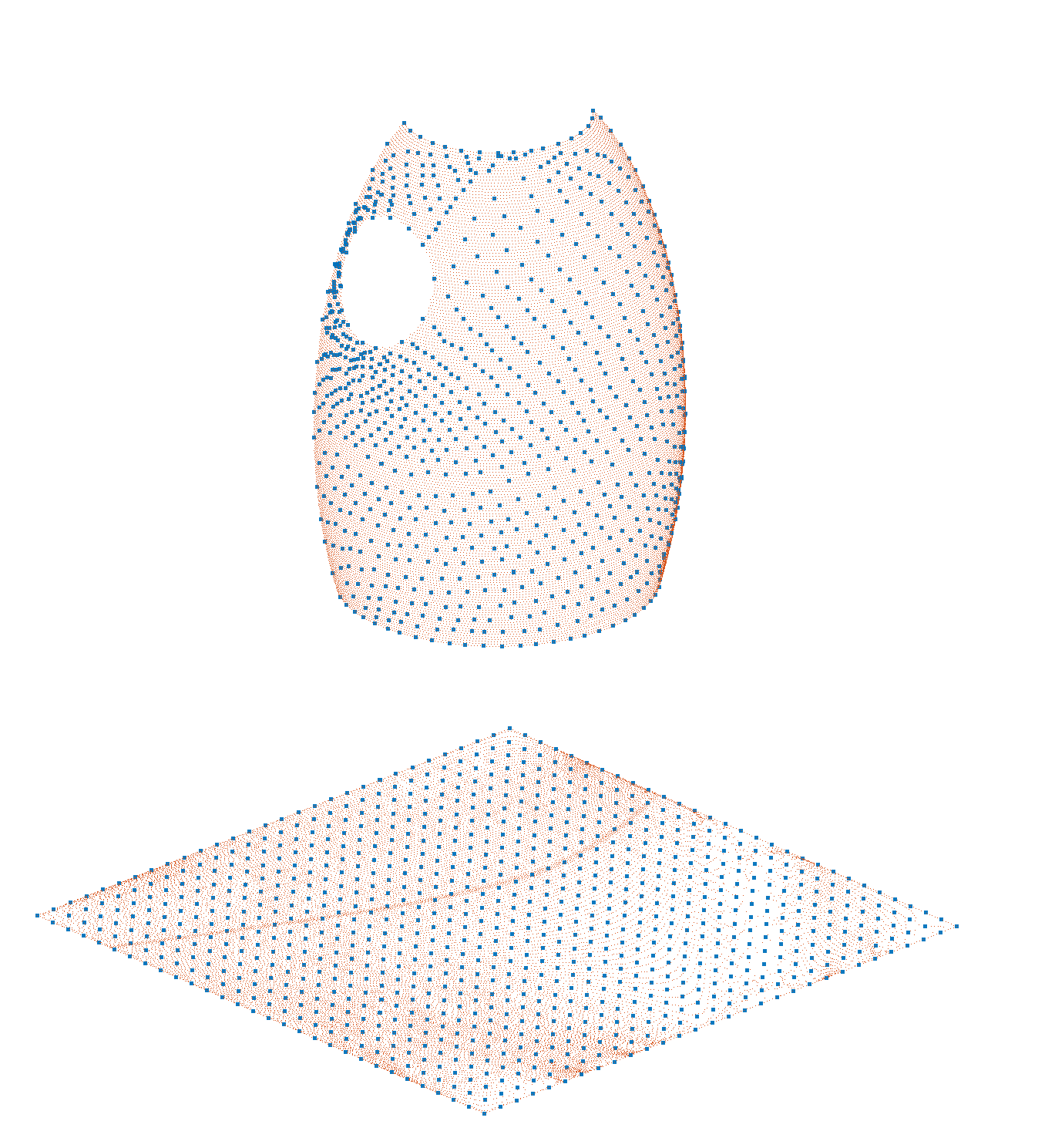}
	\caption{Parametrization by a rectangle of the rightmost $2$-cell of Figure \ref{f:armilla}. The bounding $1$-cells are mapped isometrically to a flat rectangle and then interior points are obtained using the neighboring relationships of the cloud. The $2$-cell consisting of 27 000 points (shown in red in Figure \ref{f:armilla}) is down-sampled interpolating linearly to a cloud of 900 points (shown here in blue in the parametrizing rectangle and in the $2$-cell).} \label{f:armilla_param}
\end{figure}


In Figure \ref{f:armilla_param} we show a parametrization by a rectangle of one of the $2$-cells of the vest (the red one on the right in Figure \ref{f:armilla}). This is done as explained in Section \ref{sec:param2cells}: the bounding $1$-cells are mapped isometrically to a flat rectangle (notice that we consider the $1$-cells number 6 and 6' as different) and then interior points are obtained using the neighboring relationships of the cloud (taking care of removing neighbors at opposite sides of the $1$-cell number 6). Finally, the $2$-cell consisting of 27 000 points (shown in red in Figure \ref{f:armilla_param}) is down-sampled using the parametrization and interpolating linearly to a cloud of 900 points (shown in blue in Figure \ref{f:armilla_param}).

\subsection*{Turbine engine fan blade}

\begin{figure}[htb!]
	\centering
	\includegraphics[width=1\linewidth]{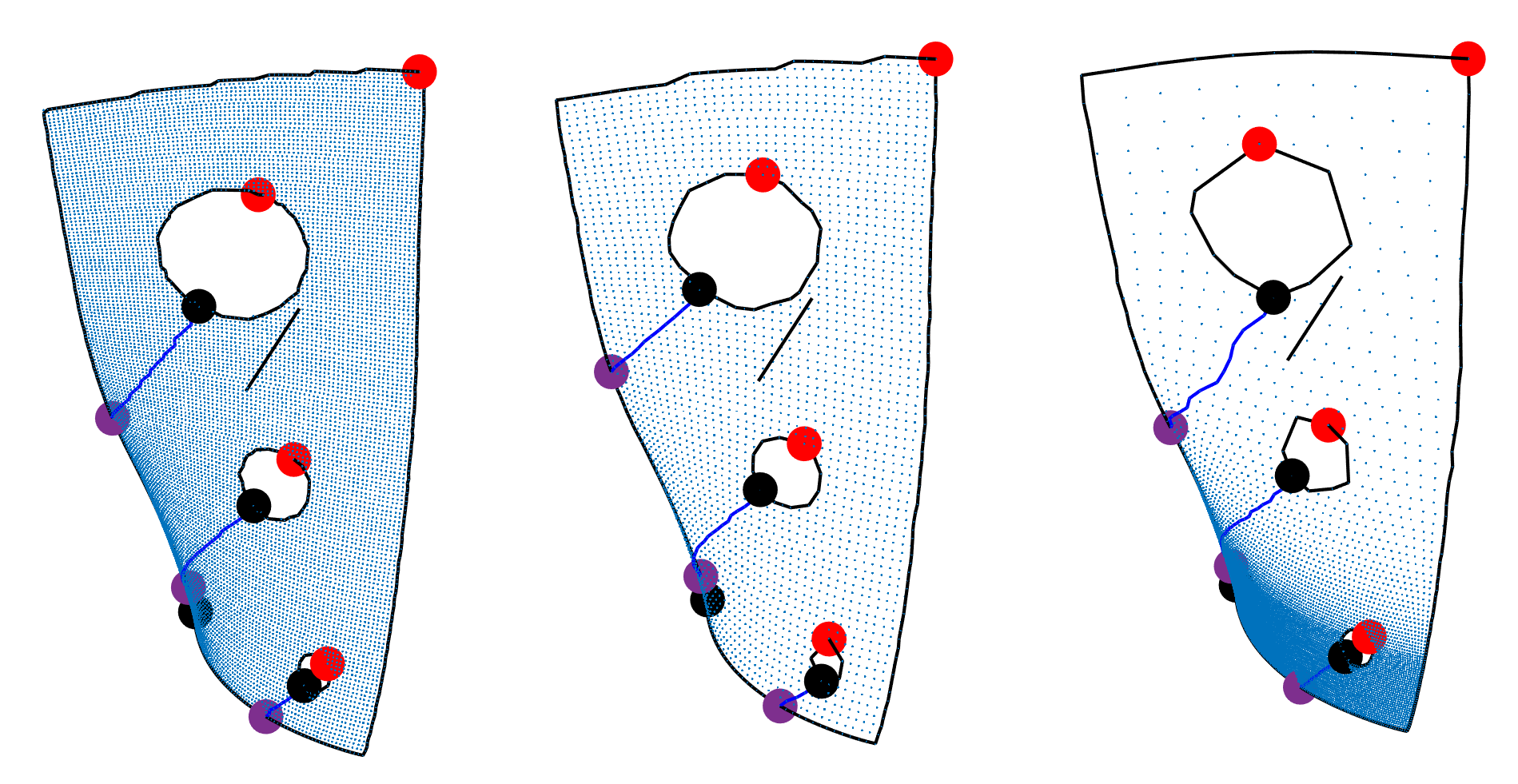}
	\caption{A 3D-scan of  an aircraft turbine engine fan blade: the black line is the direction of the height function; maxima are painted red, minima in black; $1$-cells are outlined in blue and the boundary curves in black. The purple point where the $1$-cells meet is added to the decomposition. Notice that even with 3 different sampling densities (from left to right 11000, 3000 and 7000 points) our reconstruction is robust and we get the same topological decomposition.} \label{f:turbina}
\end{figure}

Figure \ref{f:turbina} shows our algorithm applied to 3  point-clouds with different densities (approx. 3000, 7000 and 11000 points) coming from a 3D-scan of a real aircraft turbine engine fan blade, 640 mm long and 300 mm wide. For more details on how the data was collected and processed, we refer the reader to [\cite{MINEO201981}].

After detecting and parametrizing the boundary successfully, the algorithm correctly detects 1 local maxima, 3 boundary maxima, 1 local minima and 3 boundary minima for the height function depicted in the figure (the same for all 3 samples). All critical points are located at the boundary. Moreover, 3 new points (shown in purple) are added where the $1$-cells meet each other or the boundary curves (see Remark \ref{rmk:0celdas}). Then, a decomposition of the surface into 12 Morse cells is found (seven $0$-cells, ten $1$-cells and one $2$-cell). With these example clouds we highlight the robustness of our algorithm when it faces non-uniform distributions of data points.

\begin{figure}[H]
	\centering
	\includegraphics[width=0.8\linewidth]{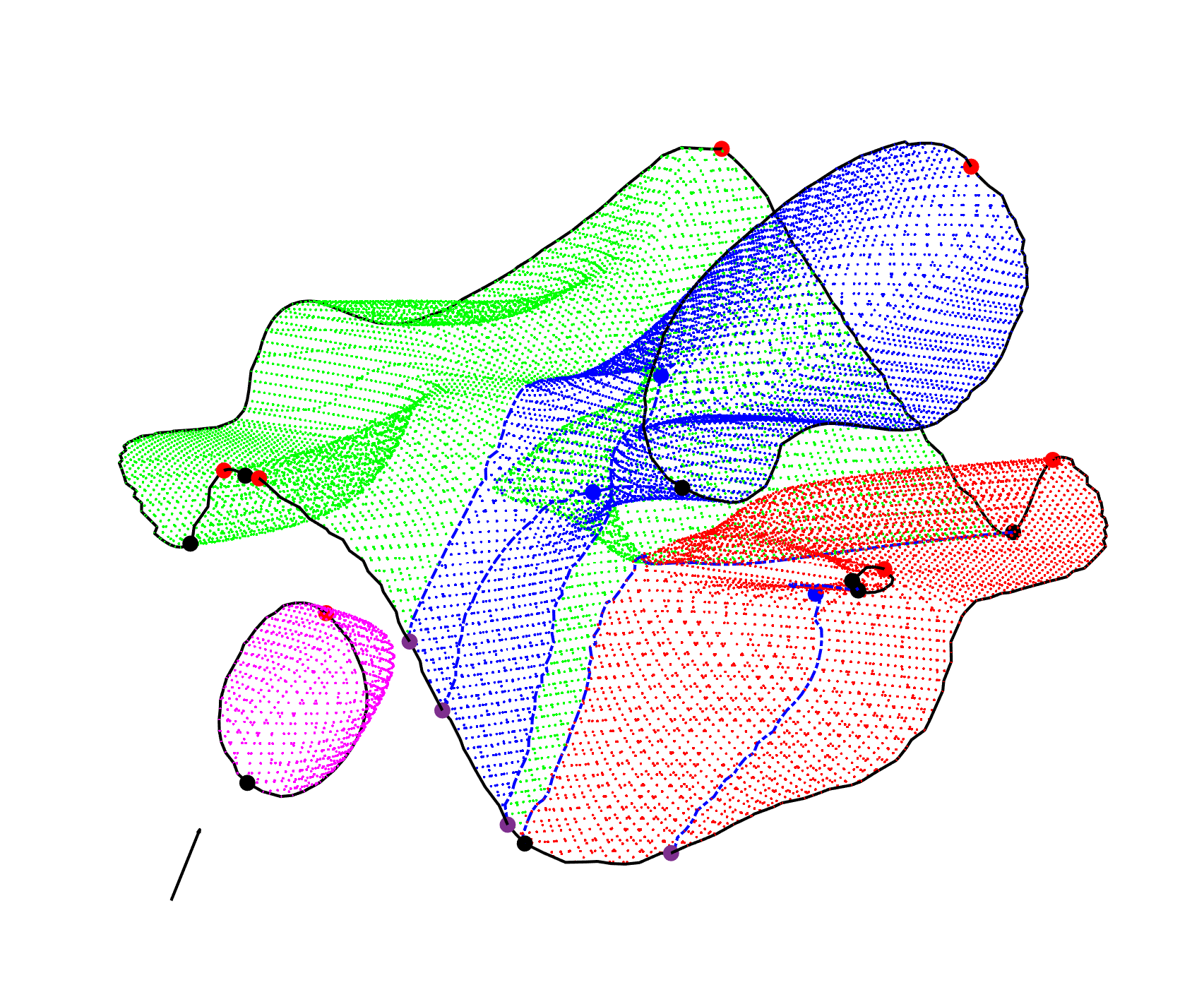}
	\caption{A sampled algebraic surface: the black line is the direction of the height function; maxima are painted red, minima in black; $1$-cells are outlined in blue and the boundary curves in black. The purple point where the $1$-cells meet is added to the decomposition.} \label{f:roseta}
\end{figure}

\subsection*{Algebraic rosette}

Figure \ref{f:roseta} shows our algorithm applied to a point-cloud coming from the zeros of a random and sparse polynomial of degree $8$ in 3 variables. This means that each point $(x,y,z)$ of the cloud satisfies an implicit equation of the form $\sum_{a,b,c = 0}^8q_{abc}x^ay^bz^c = 0$ where most of the coefficients $q_{abc}\in\mathbb{Q}$ are zero and chosen at random for $a,b,c\in\{0,1,\dots,8\}$. After detecting and parametrizing the boundary successfully, the algorithm correctly finds a decomposition of the surface into 32 Morse cells: twelve $0$-cells, sixteen $1$-cells and four $2$-cells. See the Supplementary Video 2 for a 3D view of the decomposition were all Morse cells and critical points can be appreciated.

\subsection*{3D-scan of pants}
\begin{figure}[htb]
	\centering
	\includegraphics[width=0.9\linewidth]{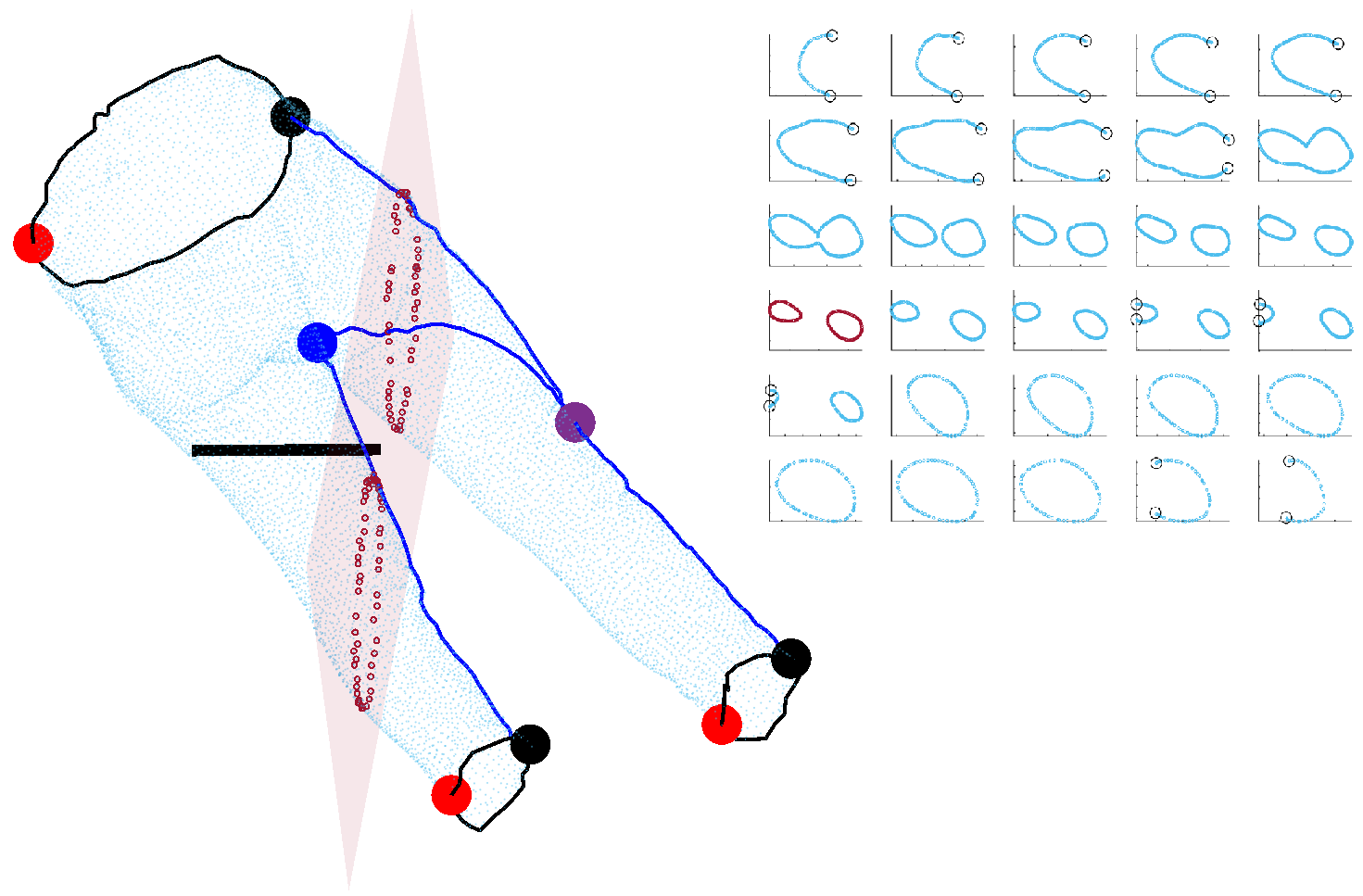}
	\caption{A 3D-scan of real pants: the black line is the direction of the height function; maxima are painted red, minima in black, saddle points are painted blue; $1$-cells are outlined in blue and the boundary curves in black. The purple point where the $1$-cells meet is added to the decomposition. On the bottom we plot the level-set curves obtained by intersecting the surface with planes perpendicular to the height function, with one such example curve shown in red.} \label{f:pants}
\end{figure}

Figure \ref{f:pants} illustrates our algorithm applied to a 3D scan of a pair of real jeans. The scan was conducted using an \textit{Artec Eva} professional handheld 3D scanner while a person was wearing the garment. 

The 3D scan was processed with the proprietary software \textit{Artec Studio} to generate a point-cloud sample. To apply our algorithm and reduce irrelevant details and noise, we down-sampled the initial cloud of over 500,000 points to 11,000 using a \textit{box-grid filter}. This filter works by computing an axis-aligned bounding box for the entire point-cloud and dividing it into grid boxes. The points within each grid box are merged by averaging their positions.

\smallskip

Despite the down-sampling, the cloud still retains significant detail (e.g., wrinkles), which increases the number of critical points and consequently, the number of Morse cells. After calculating the neighbors for each point as described in Section \ref{sec:vecinos}, we apply the \textit{discrete-curvature filter} from Section \ref{sec:filtro}. As detailed there, this involves replacing each point $v$ with a weighted average of its position and that of its neighbors (using $\alpha = 0.4$ and applying the filter 8 times). This filter establishes a bijection between the original cloud and the filtered one, preserving the topology of the underlying surface. Therefore, the decomposition we obtain from the filtered case remains valid and topologically accurate for the original cloud, as shown in Figure \ref{f:pants}.

\smallskip

The algorithm successfully detects 1 local maximum, 2 boundary maxima, 2 local minima, 1 boundary minimum, and 1 saddle point for the height function depicted in Figure \ref{f:pants}. A decomposition of the surface into 9 Morse cells is obtained: four $0$-cells (the minima and an additional point due to the intersection of original $1$-cells), six $1$-cells (associated with the saddle points, boundary minima, and boundary curves), and one $2$-cell corresponding to the local maximum. At the bottom of Figure \ref{f:pants}, we display the level set curves $\varGamma(c) = G_{\text{down}} \cap H_{c}$ (see Section \ref{sec:plane_sections}). Notice that for this point-cloud, several of the potential local transformations of level-set curves for surfaces with boundaries, as shown in Figure \ref{casos_borde}, are encountered.

\section{Conclusions and further work}
In this work, we addressed the problem of reconstructing surfaces with unknown topology from point-clouds. We introduced an algorithm capable of reconstructing a surface $S$ by obtaining a Morse cellular decomposition of it from a set of sampled points. The algorithm can be applied to surfaces in $\mathbb{R}^N$, for any ambient dimension $N$, whether the surface has a boundary or not. We demonstrated the versatility of our approach by reconstructing several surfaces, both with and without boundaries, each presenting distinct challenges. For instance, the \textit{torus} was really slim and its embedding described a (2,3)-toric knot which caused far away parts of the surface (as measured by geodesic distance) to be really near each other in Euclidean space. The \textit{rosette} is an algebraic surface which has two distinct connected components, very large and very tiny boundary curves and moreover nontrivial curvature everywhere which causes the appearance of many critical points. On the other hand, the \textit{pants} were obtained as a 3D-scan of a real pair of jeans and thus its point-cloud presented wrinkles, noise and an irregular density distribution of points. For all surfaces a global piecewise decomposition was found, with a small number of pieces in the examples examined. From the cellular decomposition the topology of the surface was computed easily (see Table \ref{tab:topology}). { Moreover, the Morse-Smale complex defined by the cells provides loops realizing any homology class, with loop length which has the order of magnitude of the shortest loop in the homology class.} The algorithm is robust: it always produces a surface, and it captures the topological features of the sampled surface with a size greater than the average distance between sample points. To obtain this decomposition, we presented a novel method to determine how (discrete) Morse cells attach to each order (see Figure \ref{f:armilla}). Furthermore, for the case of surfaces with boundary, we developed a novel graph theoretical method to determine robustly boundary points of the cloud. 

\smallskip

Further work involves studying more deeply the parametrization of the $2$-cells by flat regions of the plane. If a given $2$-cell is too far away from being flat (e.g. because of high curvature) the parametrization method described in Section \ref{sec:param2cells} may produce interior points in the region with non-uniform densities and thus it may be better to subdivide the cell into smaller and flatter pieces and afterwards parametrize each piece independently. On the other hand, as already commented in Remark \ref{r:param}, it may be interesting to use flat regions whose boundary is a polygon where its sides are mapped isometrically to each of the bounding $1$-cells, which in turn has the advantage of allowing the gluing of the resulting parametrizations of the $2$-cells into a global piece-wise defined and continuous parametrization of the entire surface. 

\smallskip
We also plan to test the reconstruction algorithm with more \textit{real} point-clouds of various textiles (e.g., those obtained via a 3D scanner). This would enable the creation of a reconstructed textile database that could later be simulated using a physical cloth model. In addition, we aim to extend the point-cloud reconstruction algorithm for surfaces to higher-dimensional manifolds by iterating the hyperplane sections, reconstructing the manifold from lower-dimensional slices. A compelling example of this would be the study of real algebraic varieties of any dimension, where point-cloud samples can be derived from their equations and refined as needed. We hope this will allow our method to detect a Whitney stratification, leading to the Morse cellular decomposition of the variety, as in [\cite{Goresky1988StratifiedMT}], through entirely numerical means.
\section{Acknowledgments}

This work was developed in the context of the project CLOTHILDE (``CLOTH manIpulation Learning from DEmonstrations") which has received funding from the European Research Council (ERC) under the European Union's Horizon 2020 research and innovation programme (grant agreement No. 741930). F.
Coltraro is supported by CSIC project 202350E080 and SGR RobIRI 2021
SGR 00514. M. Alberich-Carrami\~nana { and J. Amor\'os are} also with the Barcelona Graduate School of Mathematics (BGSMath) and the Institut de Matem\`atiques de la UPC-BarcelonaTech (IMTech), and partially supported by the Spanish State Research Agency
AEI/10.13039/501100011033 grant PID2023-146936NB-I00 
and by the AGAUR project 2021 SGR 00603 Geometry of Manifolds and Applications, GEOMVAP.

\bibliographystyle{myplain}
\bibliography{Bibliography}   

\begin{thebibliography}{10}

\bibitem{MorseEACA}
M. Alberich-Carrami\~{n}ana, J. Amor\'os, F. Coltraro, C. Torras, and M.
  Verdaguer.
\newblock Morse cell decomposition and parametrization of surfaces from
  point-clouds.
\newblock {\em Proceedings of XVII EACA 2022 (Encuentro \'Algebra Computacional
  y Aplicaciones)}, pages 35--38, 2022.

\bibitem{NormalEstimation}
J. Cao, H. Chen, J. Zhang, Y. Li, X. Liu, and C. Zou.
\newblock Normal estimation via shifted neighborhood for point cloud.
\newblock {\em Journal of Computational and Applied Mathematics}, 329:57--67,
  2018.
\newblock The International Conference on Information and Computational
  Science, 2--6 August 2016, Dalian, China.

\bibitem{Cazals2013TowardsMT}
F. Cazals, A. Roth, C.~H. Robert, and M. Christian.
\newblock Towards morse theory for point cloud data.
\newblock {\em Research Report RR-8331, INRIA.}, page~37, 2013.

\bibitem{Colome2018}
A. {Colom\'e} and C. {Torras}.
\newblock Dimensionality reduction for dynamic movement primitives and
  application to bimanual manipulation of clothes.
\newblock {\em IEEE Transactions on Robotics}, 34(3):602--615, 2018.

\bibitem{Colome2020}
A. Colom\'e and C. Torras.
\newblock {\em Reinforcement Learning of Bimanual Robot Skills.}
\newblock Volume 134 of Springer Tracts in Advanced Robotics., 2020.

\bibitem{MorseCEIG2}
F. Coltraro, J. Amor\'os, M. Alberich-Carrami\~{n}ana, and C. Torras.
\newblock {Reconstruction of sampled surfaces with boundary via Morse theory}.
\newblock In J. Gimeno~Sancho and M. Comino~Trinidad, editors, {\em Spanish
  Computer Graphics Conference (CEIG)}. The Eurographics Association, 2023.

\bibitem{CraneCurvatureFlow}
K. Crane, U. Pinkall, and P. Schr\"{o}der.
\newblock Robust fairing via conformal curvature flow.
\newblock {\em ACM Trans. Graph.}, 32(4), jul 2013.

\bibitem{Dey2006CurveAS}
T.~K. Dey.
\newblock {\em Curve and Surface Reconstruction: Algorithms with Mathematical
  Analysis}.
\newblock Cambridge Monographs on Applied and Computational Mathematics, 2006.

\bibitem{Doumanoglou2016}
A. Doumanoglou, J. Stria, G. Peleka, I. Mariolis, V. Petrik, A. Kargakos, L.
  Wagner, V. Hlavac, T.-K. Kim, and S. Malassiotis.
\newblock Folding clothes autonomously: A complete pipeline.
\newblock {\em {IEEE} Transactions on Robotics}, 32(6):1461--1478, 2016.

\bibitem{Floater2002ParameterizationOT}
M.~S. Floater and K. Hormann.
\newblock Parameterization of triangulations and unorganized points.
\newblock In {\em Tutorials on Multiresolution in Geometric Modelling}, 2002.

\bibitem{Gao2008MorseSmaleD}
J. Gao, R. Sarkar, and X. Zhu.
\newblock Morse-smale decomposition , cut locus and applications in sensor
  networks.
\newblock {\em Pre-print}, 2008.

\bibitem{Garcia-Camacho2020}
I. {Garcia-Camacho}, M. {Lippi}, M.~C. {Welle}, H. {Yin}, R. {Antonova}, A.
  {Varava}, J. {Borr\`as}, C. {Torras}, A. {Marino}, G. {Aleny\`a}, and D.
  {Kragic}.
\newblock Benchmarking bimanual cloth manipulation.
\newblock {\em IEEE Robotics and Automation Letters}, 5(2):1111--1118, 2020.

\bibitem{gensane2004dense}
T. Gensane.
\newblock Dense packings of equal spheres in a cube.
\newblock {\em The Electronic Journal of Combinatorics}, pages R33--R33, 2004.

\bibitem{Goresky1988StratifiedMT}
M. Goresky and R. MacPherson.
\newblock {\em Stratified morse theory}.
\newblock Springer Verlag, 1988.

\bibitem{Hirsch1976DiffTopo}
M. Hirsch.
\newblock {\em Differential Topology}.
\newblock Springer Verlag, 1976.

\bibitem{Hoppe1992}
H. Hoppe, T. DeRose, T. Duchamp, J. McDonald, and W. Stuetzle.
\newblock Surface reconstruction from unorganized points.
\newblock In {\em Proceedings of the 19th annual conference on computer
  graphics and interactive techniques}, pages 71--78, 1992.

\bibitem{huang2022surface}
Z. Huang, Y. Wen, Z. Wang, J. Ren, and K. Jia.
\newblock Surface reconstruction from point clouds: A survey and a benchmark,
  2022.

\bibitem{jangir2020dynamic}
R. Jangir, G. Alenya, and C. Torras.
\newblock Dynamic cloth manipulation with deep reinforcement learning.
\newblock {\em IEEE International Conference on Robotics and Automation (ICRA),
  Paris, France}, pages 4630--4636, 2020.

\bibitem{Li2015}
Y. Li, Y. Yue, D. Xu, E. Grinspun, and P. Allen.
\newblock Folding deformable objects using predictive simulation and trajectory
  optimization.
\newblock {\em Proceedings IEEE/RSJ International Conference on Intelligent
  Robots and Systems}, pages 6000--6006, 12 2015.

\bibitem{maley2005areas}
F.~M. Maley, D.~P. Robbins, and J. Roskies.
\newblock On the areas of cyclic and semicyclic polygons.
\newblock {\em Advances in Applied Mathematics}, 34(4):669--689, 2005.

\bibitem{MINEO201981}
C. Mineo, S.~G. Pierce, and R. Summan.
\newblock Novel algorithms for 3d surface point cloud boundary detection and
  edge reconstruction.
\newblock {\em Journal of Computational Design and Engineering}, 6(1):81--91,
  2019.

\bibitem{Munkres1984ElementsOA}
J.~R. Munkres.
\newblock {\em Elements of algebraic topology}.
\newblock Addison-Wesley, 1984.

\bibitem{Nealen:2006:PDM}
A. Nealen, M. M\"{u}ller, R. Keiser, E. Boxerman, and M. Carlson.
\newblock Physically based deformable models in computer graphics.
\newblock {\em Computer Graphics Forum}, 25(4):809--836, 2006.

\bibitem{Corrales2018}
J. Sanchez, J.-A. Corrales, B.-C. Bouzgarrou, and Y. Mezouar.
\newblock Robotic manipulation and sensing of deformable objects in domestic
  and industrial applications: a survey.
\newblock {\em The International Journal of Robotics Research}, 37(7):688--716,
  2018.

\bibitem{Taylor:2005:CM}
J.~R. Taylor.
\newblock {\em Classical Mechanics}.
\newblock University Science Books, 2005.

\bibitem{yin2021modeling}
H. Yin, A. Varava, and D. Kragic.
\newblock Modeling, learning, perception, and control methods for deformable
  object manipulation.
\newblock {\em Science Robotics}, 6(54):eabd8803, 2021.

\bibitem{Zhu2009TopologicalDP}
X. Zhu, R. Sarkar, and J. Gao.
\newblock Topological data processing for distributed sensor networks with
  morse-smale decomposition.
\newblock {\em IEEE INFOCOM 2009}, pages 2911--2915, 2009.

\end{thebibliography}

\appendix

\section{Morse theory for surfaces with boundary}

In this appendix we give more details on how Morse theory is applied to surfaces with boundary following the notions introduced in Section \ref{sec:morse_theory_bnd} and the tools developed in [\cite{Goresky1988StratifiedMT}]. Since the boundary of a surface with boundary is itself a manifold, we will adopt a \textit{Whitney stratification} dividing the surface into two \textbf{strata}: the first $E_1 = \partial S$ being the boundary and the second $E_2 = \text{int}(S)$ the interior. Each stratum will have as before Morse data, but this time apart from the sets $(A(p),B(p))$ defined as in Definition \ref{morse_data}, which we will call \textit{tangential data}, we will have also \textit{normal data}. 

\begin{definition}[Tangential and normal data]
	Let $p\in S$ be a critical point for some stratum $E_i$ of the Morse function $f|_{E_i}$. Then
	
	\begin{enumerate}
		\item The tangential Morse data are the pair of sets $\left(A_T,B_T\right)$ defined as in Definition \ref{morse_data} for $f|_{E_i}$. 
		
		\item The normal Morse data are the sets $\left(A_N,B_N\right)$ given by
		\begin{align*}
			A_N(p) &:= \mathcal{N}(p) \cap f^{-1}\left([f(p)-\epsilon,f(p)+\epsilon]\right),\\
			B_N(p) &:= \mathcal{N}(p) \cap f^{-1}\left(f(p)-\epsilon\right).
		\end{align*}
		where $\mathcal{N}(p) = S\cap D(p)$ is called the \textit{normal slice} at $p$ and $D(p)\subset\mathbb{R}^N$ is a sufficiently small closed disk around $p$ of dimension $N-\dim(E_i)$ transversal to $E_i$ at $p$ and the value of $\epsilon > 0$ is such that there are no more critical points in $f^{-1}[f(p)-\epsilon,f(p)+\epsilon]$. 
	\end{enumerate}
	Notice that by construction $E_i \cap D(p) = \{p\}$, and as before $B_{N,T}\subseteq\partial A_{N,T}$.
\end{definition}
\remark When $p\in\text{int}(S)$ and $N = 3$ then $D(p)$ is homeomorphic to a piece of curve normal to $S$ at $p$ and hence $\mathcal{N}(p) = \{p\}$. Therefore $(A_N,B_N) = (p,\emptyset)$. It is not hard to see that this is also the case when $N>3$.


We are now ready to state the main theorem of stratified Morse theory [\cite{Goresky1988StratifiedMT}] (SMT theorem, pages 6-8).

\begin{theorem}[Goresky-MacPherson]
	As $c\in\mathbb{R}$ increases two things can happen:
	\begin{description}
		\item[A] If between $c_1 < c_2$ there are no critical points of $f$, then $f_{\leq c_1}$ and $f_{\leq c_2}$ are diffeomorphic.
		\item[B] If there is a critical point $p\in S$ such that $c_1 < f(p) < c_2$, then $f_{\leq c_2}$ is obtained from $f_{\leq c_1}$ by performing a \textit{surgery} around $p$ with Morse data $(A,B)$ diffeomorphic to the topological product
		\begin{equation*}
			(A_N,B_N)\times (A_T,B_T) = (A_N\times A_T,A_N\times B_T\cup B_N\times A_T),
		\end{equation*}
		i.e. $f_{\leq c_2}\simeq \left(f_{\leq c_1}\cup A\right)\slash\sim$ where the equivalence relation is given by identifying $B\subseteq f_{\leq c_1}$ with points of $\partial A\supseteq B$.  
	\end{description}
\end{theorem}

\remark We already established that at interior critical points of $S$, we have $(A_N,B_N) = (p,\emptyset)$. Therefore in that case the above product is trivial and the attachment maps are the same ones described earlier. 

\begin{figure}[H]
	\centering
	\includegraphics[scale=.25]{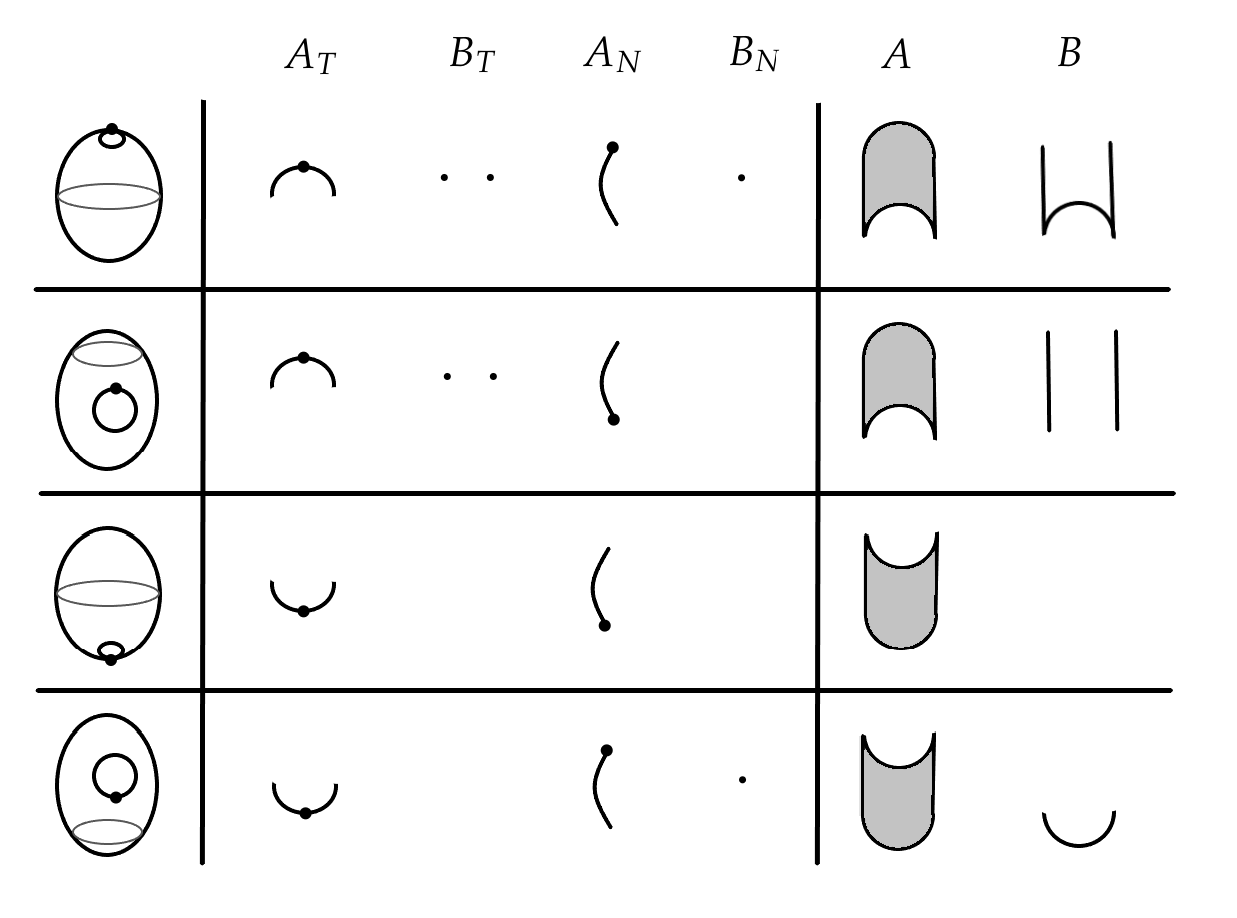}
	\caption{Four types of critical points located at the boundary and their tangential and normal Morse data.}
	\label{cirugias_borde}
\end{figure}

The new cases occur when $p$ is a critical point of $f|_{\partial S}$ lying on $\partial S$. Since $\partial S$ is a one-dimensional curve, $p\in \partial S$ can only be a maximum or a minimum. Nevertheless, depending on whether $p$ is also a local minima (resp. maxima) of $f$ or just of $g = f|_{\partial S}$, we will have four different cases (see Figure \ref{cirugias_borde}). Recall that for minima (and maxima) points $p$ of $f$ located at $\partial S$ we have that $\text{d}_pf \neq 0$. Nevertheless, in order not to complicate the discussion semantically we will still call them critical points since they satisfy $\text{d}_pg = 0$.


\remark We will denote by $\left(\blacksquare,\sqcup,|\;|,\_\right)$ a quadrangular $2$-cell, all its sides minus the top one, two lateral sides and the bottom side, respectively.

The four new types of critical points that we can have are:

\begin{description}
	\item \textit{Maxima} of $f|_{\partial S}$: In this case $A_T$ is a closed concave piece of curve (which we denote by $\cap$) and $B_T$ two points, i.e. $(A_T,B_T) = \left(\cap,.\;.\right)$. We now have two sub-cases:
	\begin{enumerate}
		\item $p$ is also a local maximum of $f$. Then $A_N$ is a closed interval and $B_T$ one point, i.e. $(A_N,B_N) = \left(\;|\;,.\;\right)$. Therefore the topological product is
		\begin{equation*}
			(A,B) = \left(\cap,.\;.\right)\times \left(\;|\;,.\;\right) = \left(\blacksquare,\sqcup\right),
		\end{equation*}
		and the three of the sides of $\partial A$ where $p$ is not present are attached to $B$. During this surgery a $1$-cell (containing $p$) is attached to the boundary and a $2$-cell to the interior (closing a void in the process). 
		\item $p$ is not a local maximum of $f$ (only of $f|_{\partial S}$). Then $A_N$ is again a closed interval but $B_T$ is empty, i.e. $(A_N,B_N) = \left(\;|\;,\emptyset\right)$. Then the Morse data is 
		\begin{equation*}
			(A,B) = \left(\cap,.\;.\right)\times \left(\;|\;,\emptyset\right) = \left(\blacksquare,\;|\;|\;\right),
		\end{equation*}
		and the two opposite sides of $\partial A$ where $p$ is not present, attach to $B$. This surgery is equivalent to attaching a $1$-cell to the boundary (but no $2$-cell is attached to the interior).
	\end{enumerate}
	\item \textit{Minima} of $f|_{\partial S}$: In this case $A_T$ is a closed convex piece of curve (which we denote by $\cup$) and $B_T$ is empty, i.e. $(A_T,B_T) = \left(\cup,\emptyset\right)$. We again have two sub-cases:
	\begin{enumerate}
		\item $p$ is also a local minimum of $f$. Then $A_N$ is a closed interval and $B_T$ is empty, i.e. $(A_N,B_N) = \left(\;|\;,\emptyset\right)$. Therefore the topological product is 
		\begin{equation*}
			(A,B) = \left(\cup,\emptyset\right)\times \left(\;|\;,\emptyset\right) = \left(\blacksquare,\emptyset\right).
		\end{equation*}
		In this case there is no surgery, the cell $A$ just appears. This cell retracts to a point, which will be a $0$-cell in the cell complex.
		\item $p$ is not a local minimum of $f$ (only of $f|_{\partial S}$). Then $A_N$ is a closed interval and $B_T$ is a point, i.e. $(A_N,B_N) = \left(\;|\;,.\;\right)$.  The Morse data is 
		\begin{equation*}
			(A,B) = \left(\cup,\emptyset\right)\times \left(\;|\;,.\;\right) = \left(\blacksquare,\_\right).
		\end{equation*}
		and the opposite side of $\partial A$ where $p$ is, attaches to $B$. This surgery is equivalent to attaching a $0$-cell to the boundary and a $1$-cell to the interior.
	\end{enumerate}
\end{description}

\end{document}